Technical Report

# An Efficient Solution for Model Checking Abstract State Machine Using Bogor

**Saeed Doostali**

Department of Computer Engineering, Faculty of Engineering, Arak University, Arak, Iran

**Abstract:** Nowadays, publish-subscribe (pub-sub) and event-based architecture are frequently used for developing loosely coupled distributed systems. Hence, it is desirable to find a proper solution to specify different systems through these architectures. Abstract state machine (ASM) is a useful means to visually and formally model pub-sub and event-based architectures. However, modeling per se is not enough since the designers want to be able to verify the designed models. As the model checking is a proper approach to verify software and hardware systems. In this paper, we present an approach to verify ASM models specified in terms of AsmetaL language using Bogor – a well known model checker. In our approach, the AsmetaL specification is automatically encoded to BIR, the input language of the Bogor.

## Contents

## 1. Introduction

Nowadays, using of pub-sub and event-based architectures for designing large scale distributed systems, including real-time systems, management and control systems, and electronic commerce is prevalent [1]. These architectures provide a coordination model to integrate components in loosely coupled systems. On the other hands, the event-based components are not designed to work with specific other components. It facilitates the integration of autonomous and heterogeneous components into complex applications. So, these architectures provide the flexibility, scalability and adaptability that is required in many application domains.

Abstract state machine (ASM) [2, 3] is a formal method which is frequently used to design pub-sub and event-based systems. Basic concepts of ASMs were introduced by Yuri Gurevich in 1980. ASMs are an extension of finite state machines. They are used for variety of domains e.g., embedded systems, protocols, formal specification and analysis of computer hardware and software, specification of programming languages (like C and JAVA), and design languages (like UML). ASM theory was expressed as: "any algorithm, no matter how abstract, is step-for-step emulated by an appropriate ASM" [4], and it is the main characteristic of the ASM methods, e.g., it allows the designer to specify a system at a natural level of abstraction.

However, the modeling per se is not enough and verification is an important issue. In many of the systems, especially safety critical ones, existing subtle errors may cause loos of human resources. It can be very difficult to discover these errors. Among different existing approaches for the verification, model checking is one of the most accurate solutions. Model checking is an efficient and automated formal verification method for evaluating a system.

ASM is a useful means to formally model event-based and pub-sub systems. There are different notations and language along with different toolset to state ASM specification. Moreover, each tool introduces a different syntax for representing the ASM models, and provides special constructs. So, to integrate these different tools and notations the OMG meta-modeling framework, called ASMETA (ASM mETAmodeling)[1] [5] has been proposed. It also provides a standard framework for developing new ASM tools. The ASMETA toolset consists of tools for the creation, editing, simulation [5], validation [6, 7], verification [8], and runtime monitoring [9, 10] of ASM models.

In this paper, we present an approach to verify systems specified through AsmetaL using model checking. This paper is an extended version of the work presented in [11]. We introduce the architecture of the implemented tool ASM2Bogor along with the proposed algorithms. To do so, we encode AsmetaL specifications to BIR along with the desired properties to be verified. Then, Bogor [12] generates the transition system and checks the stated properties on the transition system. The encoding procedure is done automatically, so the designers do not need to know the syntax of Bogor.

The rest of the paper is organized as follows: Section 2 presents the state of the art. In Section 3 and 4, we briefly introduce the required background, i.e., the model checker Bogor, and ASMETA and the AsmetaL Language. The architecture of our translator is presented in Section 5. In Section 6, we describe our encoding approach to encode AsmetaL to BIR. In Section 7, we describe, with some case studies, how we can use of ASM2Bogor to model check ASM models by means of the model checker Bogor. Finally, section 8 concludes the paper.

---

[1] ASMETA website (http://asmeta.sourceforge.net/)

## 2. Related work

Winter presents an approach for transforming ASMs into the input of the model checker SMV (Symbolic Model Verifier) [13]. Later on, Winter and Del Castillo improve the approach for translating the ASM-SL specifications into the SMV [14]. Winter also presents an extension to the approach to verify ASMs using the model checker MDG (Multiway Decision Graph) [15]. They apply their approach to a complex case study in [16]. In their approach, ASMs must be flattened into set of conditional rule with only nullary functions, called location unfolding. So, the ASMs that can be used are very limited and their translator, ASM2SMV, is no longer available.

In [17], Gargantini et al. present an approach to encode an ASM in Promela, the input language of model checker Spin. The transforming approach does not support n-ary functions and extended rule forms like forall rules. Their approach is significantly improved by Farahbod et al. [18]. They translate ASM models written in CoreAsm [19], into Promela and use Spin to verify properties of CoreAsm specifications. Up to now, a tool which supports the approach has not been released.

Spielmann [20] provides an algorithm for verifying ASM by means of a Computation Graphs Logic (CGL*) which is similar to the CTL* temporal logic. He proposes to translate an ASM model into NuSMV for checking properties specified through CGL*. Although the approach can check the ASMs with infinitely many inputs, however, it is severely limits the ASM models since it supports only nullary functions.

In another approach, Kardos develop a model checker for AsmL language [21], in [22]. The approach directly explores the state space to verify an AsmL specification. Kardos believes that using of the existing model checker limits the supported ASMs. Beckers et al. [23] also present an approach for verifying ASMs without translating ASMs into the input language of the existing model checkers. They use of CoreAsm to generate the state space and combine it with the model checker [mc] square. The benefit of such idea is that the input language can be very expressive. Nonetheless, their approach cannot be applied on the complex specification and they are very inefficient, because it is not possible to do all the optimizations that other model checkers (like Spin and Bogor) can do them.

More related to our approach, Arcaini et al. [8] propose an approach for model checking ASMs. They translate ASMs which are written in AsmetaL into the input language of the model checker NuSMV to verify different properties. They try to provide a method that covers a broader range of ASM models, but they do not support all the AsmetaL constructs (like turbo rules). Moreover, their approach also needs to location unfolding.

## 3. The model checker Bogor

Bogor is an extensible and customizable model checker for verification of dynamic and concurrent systems [3]. It provides a framework for extending the modeling language called BIR (Bandera Intermediate Representation) aimed at designing domain-specific systems with new data types, commands, and expressions. In fact, Bogor reduces the semantic gap between the input language and the system description which exists in other model checkers such as Spin [24] and NuSMV [25].

BIR is the input language of Bogor which resembles Promela. It supports many features of a rich base modeling language, e.g., dynamic creation of threads and objects, automatic memory management, virtual method calls, exception handling, user-defined data types and non-determinism. BIR contains two different language constructs: high-level and low-level. High-level BIR is more similar to source code of programming languages, while low-level BIR describes states and transitions in a guarded command format. An example of BIR model is shown in Fig. 1: (a) is the high-level model and (b) is its equivalent low-level model. In this model, the designer describes the Collatz conjecture as a simple loop. The first guard checks whether the global integer variable *x* is odd and the other guard checks if it is even. If the first guard is satisfied the action *x:=x/2* is executed. If none of the guards can be satisfied the thread is stopped. In this situation, Bogor detects a deadlock and creates a counter-example which allows us to see how the deadlock occurs. Notice that, all guards in an active location are evaluated simultaneously and if more than one guard is true, then Bogor non-deterministically chooses one of them. It is also possible to define different locations with the keyword **loc** and explicitly jumping among them with the instruction **goto**. For example, in Fig 1.b, the next location for the first guard is *loc0*.

```
system example {                        system example {
    int x:=100;                             int x:=100;
    main thread MAIN() {                    main thread MAIN() {
        if <x%2==0> do                          loc loc0:
            x:=x/2;                                 when x%2==0 do {x:=x/2;}
        else do                                     goto loc0;
            x:=3*x+1;                               when x%2!=0 do {x:=3*x+1;}
        end                                         goto loc0;
    }                                           }
}                        (a)             }                        (b)
```

**Fig. 1.** Example BIR model: (a) High-level, (b) low-level BIR model [26]

By default, Bogor uses the MAIN thread as a starting point of the model checking process. In the BIR model of Fig. 1, the execution flow starts by initializing *x* with 100 and checking the two guards in *loc0*. Bogor generates the transition system for verifying the BIR models, representing all the reachable states.

```
fun fail() returns boolean =
    LTL.temporalProperty(
        Property.createObservableDictionary(
            Property.createObservableKey("p", x>0),Property.createObservableKey("q", x<0)),
        LTL.always(LTL.implication(LTL.prop("p"),LTL.eventually(LTL.prop("q")))));

fun hold() returns boolean =
    LTL.temporalProperty(
        Property.createObservableDictionary(
            Property.createObservableKey("p", x>0),Property.createObservableKey("q", x<=100)),
        LTL.always(LTL.conjunction(LTL.prop("p"),LTL.prop("q"))));
```

**Fig. 2.** Two example property functions [26]

The main purpose of a model checker is to verify a set of properties. In Bogor, the specifications to be checked can be expressed in Linear Temporal Logic [27]. Bogor starts from the initial state and checks all the reachable states in the transition system.

For example, we have added two LTL specifications to the model, as shown in Fig. 2. These properties must be defined as Boolean functions in the BIR model. The equivalent LTL formula of the first property is $\square((x>0)\rightarrow\diamond(x<0))$. It is true if, whenever $x$ is greater than 0, then there is a future state (eventually) in which $x$ less than 0. Due to the BIR model of Fig. 1, this property cannot be satisfied. Fig. 3 shows the counter-example. The equivalent LTL formula of the second property is $\square(x>0 \wedge x\leq100)$. It is true if globally in each state, the value of $x$ is in the range of 0 to 100. This property is satisfied by the BIR model of Fig. 1.

| 1. x = 100 MAIN.loc0 | 2. x = 50 MAIN.loc0 | 3. x = 25 MAIN.loc0 | 4. x = 76 MAIN.loc0 | 5. x = 38 MAIN.loc0 | 6. x = 19 MAIN.loc0 | 7. x = 58 MAIN.loc0 | 8. x = 29 MAIN.loc0 |
|---|---|---|---|---|---|---|---|
| 9. x = 88 MAIN.loc0 | 10. x = 44 MAIN.loc0 | 11. x = 22 MAIN.loc0 | 12. x = 11 MAIN.loc0 | 13. x = 34 MAIN.loc0 | 14. x = 17 MAIN.loc0 | 15. x = 52 MAIN.loc0 | 16. x = 26 MAIN.loc0 |
| 17. x = 13 MAIN.loc0 | 18. x = 40 MAIN.loc0 | 19. x = 20 MAIN.loc0 | 20. x = 10 MAIN.loc0 | 21. x = 5 MAIN.loc0 | 22. x = 16 MAIN.loc0 | 23. x = 8 MAIN.loc0 | 24. x = 4 MAIN.loc0 |
| 25. x = 2 MAIN.loc0 | 26. x = 1 MAIN.loc0 | 27. repetitive state x = 4 MAIN.loc0 | | | | | |

**Fig. 3.** Bogor counter-example: state space variables

## 4. The ASM modeling language

The abstract stat machine $M$ is a tuple: $M=(S_M, T_M, I_M)$, in which $S_M$ is the set of states of $M$, $T_M$: $S_M\rightarrow S_M$ is the set of transition rules which specify the behavior of $M$, and $I_M\subseteq S_M$ is the set of initial states. The ASMETA framework provides a concrete textual notation, AsmetaL, to effectively write the ASM $M$. It is a metamodel based language.

An AsmetaL model is structured into four sections: a header, a body, a main rule and an initialization. The modeler must specify the name of the ASM model before the header section. The header section includes some imports and one export clauses to communicate this model with other ASM models or AMS modules. Domains and functions are used to model the required data structure of the ASM models. Those are declared in the signature part in the header section. AsmetaL supports several types of domains (e.g., basic domain, abstract type domain, etc) and functions (e.g., controlled, static, derived, monitored) which are useful for model different aspect of the system.

The implementation of static concrete domains, static and derived functions, rules, and axioms is inserted in the body section. The main rule which labeled by the **main** keyword is the starting point of executing an ASM. The behavior of the ASM is specified by rules in body section and main rule in AsmetaL. The rules are specified with the prefix string "r_" and divided into turbo and macro rules. The turbo rules return a value which placed to the *result* as a global variable, while the macro rules have no returned value. They consist of several transition rules which describe changing of the controlled functions from a state to another one.

Table 1. AsmetaL transition rules

| Rule | AsmetaL syntax | Comment | |
|---|---|---|---|
| Update rule | $T_1$:=v <br> $T_1$:=$T_2$ | It assigns the value $v$ to the term $T_1$ <br> It assigns the value of term $T_2$ to $T_1$ | |
| Conditional rule | **If** <cond> then <br>   $R_{if}$ <br> [**else** $R_{else}$] <br> **endif** | A conditional rule expresses branching in execution. <br> If the boolean term *cond* is satisfied then rule $R_{if}$ is executed, <br> else rule $R_{else}$ is executed. The else branch is optional. | |
| Parallel block rule | **par** <br>   $R_1$ <br>   ... <br>   $R_n$ <br> **endpar** | The grouped rules $R_1,...,R_n$ are executed in parallel. | |
| Sequential block rule | **seq** <br>   $R_1$ <br>   ... <br>   $R_n$ <br> **endseq** | The grouped rules $R_1,...,R_n$ are sequentially executed. | |
| Case rule | **switch** T <br>   **case** $v_1$: $R_1$ <br>   ... <br>   **case** $v_n$: $R_n$ <br>   [**otherwise** $R_{ow}$] <br> **endswitch** | Case rule is an extension of conditional rule. If $T=v_i$ is satisfied <br> then rule $R_i$ is executed ($1 \le i \le n$), <br> else $R_{ow}$ is executed, if otherwise branch is not null. | |
| Choose rule | **choose** $d_1$ **in** $D_1$ ,..., $d_n$ **in** $D_n$ <br> **with** <$cond_{1,...,n}$> **do** <br>   $R_{1,...,n}$ <br> [**ifnone** $R_{ifnone}$] | If there are some $V_k$ which satisfy $cond_{1,...,n}$, then choose non-deterministically selects one of them and executes the rule $R_{1,...,n}$, otherwise $R_{ifnone}$ is executed, if *ifnone* branch is not null. | $D_i$ is an enumerable collection with cardinality $|D_i|$, where $1 \le i \le n$. $cond_{1,...,n}$ is a boolean condition over the variables $d_1,...,d_n$, and $R_{1,...,n}$ and $R_{ifnone}$ are transition rules that contains occurrences of their variables. $V_k \in \{(v_1,...,v_n) \mid v_1 \in D_1,...,v_n \in D_n\}$, $v_i$ is the value of the variable $d_i$ and $k \in 1..\prod_{1 \le i \le n} |D_i|$. |
| Forall rule | **forall** $d_1$ **in** $D_1$ , ..., $d_n$ **in** $D_n$ <br> **with** <$cond_{1,...,n}$> **do** <br>   $R_{1,...,n}$ | A forall rule specifies the parallel execution of the same rule $R_{1,...,n}$ for all elements $V_k$ that satisfy $cond_{1,...,n}$. | |

Table 1 shows the syntax of the transition rules and briefly describes the statements which are supported by AsmetaL (see [28], for a complete description of AsmetaL).

Finally, the initialization section initializes each domain and dynamic functions. It also consists of agent domains with its transition rule which called program of agent.

## 5. Architecture of the tool ASM2Bogor

Fig. 4 shows the main components of the tool ASM2Bogor that translates ASM models to BIR. The designer must specify the ASM models and the properties to be checked by the AsmetaL language. We use AsmetaLc, the AsmetaL compiler, to parse the input model. Therefore, the designer only needs to know the syntax of the AsmetaL to model problems and temporal operators to express LTL specifications. Given an AsmetaL model, ASM2Bogor checks the possibility of mapping it to BIR and then generates the semantically equivalent BIR model. After the translation, ASM2Bogor directly runs Bogor for verifying the desired properties. If the property is not satisfied, Bogor generates a counter example to help the designer to correct the model.

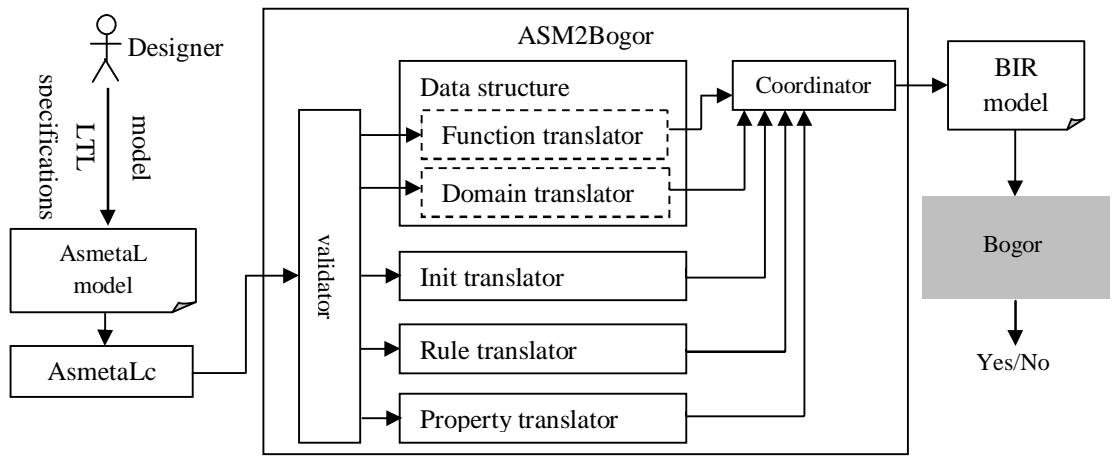

**Fig. 4.** The architecture of ASM2Bogor [11]

## 6. Mapping the ASM concept into BIR

In this section, we present an approach to automatically translate ASM models written in AsmetaL to BIR. The main steps of our approach can be summarized as follows: (a) we check the possibility of mapping the AsmetaL model to the BIR model, (b) we translate the ASM data structures by encoding the domains and functions which are defined in signature section, and are implemented in definition section, (c) we initialize them through the initial (*default init*) section, (d) finally, we encode the transitions which consist of macro and turbo rules.

### 6.1. Validation

In the first step, we decide if it is possible to translate the input model to BIR. It means that we should check the input model to find at least one structure which cannot be translated. If it is the case, ASM2Bogor cannot generate the equivalent BIR model. Otherwise, the second step is performed. For example, *forall $s in Integer with true do R* is not supported, because a forall rule over an infinite domain cannot be treated by the model checker Bogor. So, we must check domains, functions, and rules.

### 6.2. Data Structure

In the second step of our approach, we translate the necessary data structures which are required in the model. It consists of declaration and implementation of domains and functions which are used in the AsmetaL model. Their declarations are placed in the *signature* section, whereas their implementations are given in the *definition* section. We map them to the equivalent BIR elements.

### 6.2.1. Domains

An AsmetaL model that must be mapped into BIR can contain only domains that have a corresponding type in BIR. The only supported domains are described in the following.

AsmetaL supports several types of domains: (a) *basic* domain (e.g., Boolean, Real, Integer, Natural, Char, String, and Undef), (b) *abstract* domain which is useful for representing the real word entries, (c) *enumeration* domain, (d) *subset* domain, and (e) specific domain (e.g., Sequence, Set, Bag, Map, and Product domain).

We map the basic domain to the corresponding primitive data type of BIR. In order to do it, we have created the table domains where, for each basic domain, an equivalent primitive data type is declared (see Table 2). Moreover, the user defined domains must be added to the table.

We describe this mapping as a function $[\![.]\!]$, called *domain_map* function, that maps each supported AsmetaL domain into its corresponding BIR type.

**Table 2.** *Domains* table contains AsmetaL domains and their BIR domains.

| AsmetaL Domain | BIR Domain |
|----------------|------------|
| Boolean | boolean |
| Integer | int |
| Natural | int |
| String | string |
| Char | string |
| Real | float |
| Undef | null |

Abstract type domains are mapped to records with the same name. Since agents domains are particular kinds of abstract domains, they are mapped in the same way. Note that we must add the records as the user defined domains to the domains table. For an agent domain *A*, and an abstract domain *D*, the *translation rule* δ is defined as follows:

$$\delta(\textbf{domain } A \textbf{ subsetof } Agent) \Rightarrow \textbf{record } A\{\,\}$$
$$\delta(\textbf{abstract domain } D) \Rightarrow \textbf{record } D\{\,\}$$

In AsmetaL, we can define the elements of an agent and abstract domain as constant in the *signature* part. To map each of the elements, we define a variable whose type is the name of the agent or abstract domain. For example, we create a record to represent the agent domain *Philosophers*, and a record to represent the abstract domain *Forks*, and we define two variables *phil* and *fork*, as shown in Fig. 5.

```
signature:
    domain Philosophers subsetof Agent
    abstract domain Forks
    static phil: Philosophers
    static fork: Fork                    (a)
```

```
record Philosophers {}
record Forks {}
Philosophers phil;
Fork fork;                    (b)
```

**Fig. 5.** Agent and abstract domains, (a) AsmetaL notation, (b) BIR notation.

Since Bogor supports the enumerated domains directly, their transformation is straightforward. The translation rule for enumerated domain is defined as follows:

$$\delta(\textbf{enum domain } E = \{e_1|e_2|\dots|e_n\}) \Rightarrow \textbf{enum } E\{e_1, e_2, \dots, e_n\}$$

where *E* is the enumerated name and $e_1, e_2, \dots, e_n$ are the elements of the enumerated domain. For example, Fig. 6 shows the mapping of the enumerative domain *Light*.

```
signature:
    enum domain Light = {GREEN | RED}    (a)
```

```
enum Light { GREEN, RED}                    (b)
```

**Fig. 6.** Enum domain, (a) AsmetaL notation, (b) BIR notation.

To represent the subset domain, we create an alias name by *typealias* keyword. The translation rule for subset domain, $D_1$, is defined as follows:

$$\delta(\textbf{domain } D_1 \textbf{ subsetof } D_2) \Rightarrow \textbf{typealias } D_1 \; [\![D_2]\!];$$

```
asm SubsetDomain
import StandardLibrary
signature:
    domain SubInt subsetof Integer
    dynamic controlled foo: SubInt
definitions:
    domain SubInt = {1..3}
    main rule r_Main = foo:=2
default init s0:
    function foo=1
                (a)
```

```
system SubsetDomain {
    typealias SubInt int;
    SubInt foo;
    active thread MAIN() {
        loc loc0://initialization
            do {
                foo := 1;
                assert(foo>=1 && foo<=3);
            }goto loc1;
        loc loc1:
            do invisible{
                foo := 2;
                assert(foo>=1 && foo<=3);
            }goto endloc;
        loc endloc:
            do {/*Bogor creates a state*/} goto loc1;
}}                    (b)
```

**Fig. 7.** Subset domain, (a) AsmetaL notation, (b) BIR notation.

We also check the value of a function having the subset domain *D* as codomain is in its domain by an assertion when it is updated. For example, Fig. 7 shows the mapping of an integer subset domain *SubInt*.

Finally, as described in the Section 3, BIR provides the extension facility which allows us to add new data types, commands, and expressions to the basic modeling language. Hence, we define an extension for encoding the *sequence* domain. First of all, to create an extension, we must write a java package implementing the semantics of the new data types, commands, and expressions (i.e., myPackage.SeqModule). Then, we introduce the extension by an extension declaration that specifies the new symbols and associated arities. Fig. 8 shows the declaration for the sequence abstract data type.

```
extension Seq for myPackage.SeqModule{
        typedef type<'a>;
        expdef Seq.type<'a> create<'a>('a ...);
        expdef boolean isEmpty<'a>( Seq.type<'a> seq);
        expdef boolean contains<'a>( Seq.type<'a> seq, 'a elem);
        expdef int count<'a>( Seq.type<'a> seq, 'a elem);
        expdef int length<'a>( Seq.type<'a> seq);
        expdef int indexOf<'a>( Seq.type<'a> seq, 'a elem);
        expdef 'a first<'a>( Seq.type<'a> seq);
        expdef 'a last<'a>( Seq.type<'a> seq);
        expdef 'a atIndex<'a>( Seq.type<'a> seq, int pos);
        expdef Seq.type<'a> tail<'a>( Seq.type<'a> seq);
        expdef Seq.type<'a> union<'a>( Seq.type<'a> seq1, Seq.type<'a> seq2);
        expdef Seq.type<'a> subSequence<'a>( Seq.type<'a> seq, int pos1, int pos2);
        actiondef append<'a>( Seq.type<'a> seq, 'a elem);
        actiondef excluding<'a>( Seq.type<'a> seq, 'a elem);
        actiondef prepend<'a>( 'a elem , Seq.type<'a> seq);
        actiondef insertAt<'a>(Seq.type<'a> seq, int pos, 'a elem);
        actiondef replaceAt<'a>(Seq.type<'a> seq, int pos, 'a elem);
}
```

**Fig. 8.** Bogor *sequence* extension declaration.

We define a polymorphic type to represent the type of sequence elements that used by declaring a variable whose type is "*Seq.type<something>*". The *create* expression is the constructor which allows any number of elements as the initial content (i.e., *'a ...*). We also define new expressions to check a sequence for emptiness (*isEmpty*), to test a given element exists in a sequence (*contains*), to count the number of a special element in a sequence (*count*), to take the length of a sequence (*length*), to take the first index of a given element in a sequence (*indexOf*), to pick the first element from a sequence (*first*), to pick the last element from a sequence (*last*), to pick an element which is placed on a given position in a sequence (*atIndex*), to take the tail of a sequence (*tail*), to take a union of two sequence (*union*), and to pick a subsequence from a sequence (*subsequence*). Moreover, we introduce new actions that add, replace, and remove elements from a sequence. The actions *append, prepend,* and *insertAt* add an element to the end, to the first, and to a given position in a sequence, respectively. The *excluding* action removes a given element from a sequence and the *replaceAt* actions replaces an element which is placed at a given position with a new element.

The Set, Bag, Map, and Product domains are implemented in a similar way.

### 6.2.2. Functions

AsmetaL functions are divided into two sets: function with no arguments (nullary functions), and functions with *n* arguments (n-ary functions), with *n*>0.

We create the *variables* table, that contains variables and function locations which are declared in the model, and their translation into BIR. We also introduce a function $\langle\!\langle . \rangle\!\rangle$, called *Vmap* function, that maps each variable and function location into its equivalent BIR variable.

*Nullary functions:* We define constant variables for the static and derived nullary functions, because their value is fixed during the execution of the machine. For every other nullary function, we define a variable whose type is given by the mapping of the function codomain. Since Bogor introduces record variables during a run (e.g., memory space allocation), we must instantiate them with *new* keyword in *loc0*, the first location of the thread main. Note that the user defined variable must be added to the variables table.

For *n*-ary functions different categories can be considered: static and derived functions, controlled functions, and monitored functions.

*Static and derived functions:* Since the static functions express a fixed relation between the elements of the input domain and the codomain and the computation mechanism of the derived functions is fixed, we do not distinguish between the static functions and the derived functions in our approach: their mapping is the same. We define a BIR function *fun*, to encode them. The input domain determines the parameter types and the returned value type of the function. The translation rule for static/derived function, *f*, is defined as follows:

$$\delta \begin{pmatrix} \textbf{signature}: \text{static/derived} \quad f: D_1 \to D_0 \\ \textbf{definitions}: \textbf{function} \ f(d_1 \ \text{in} \ D_1) = R \end{pmatrix} \Rightarrow \ \textbf{fun} \ f(\llbracket D_1 \rrbracket \ d_1) \ \textbf{returns} \ \llbracket D_0 \rrbracket = \delta(R)$$

$$\delta \begin{pmatrix} \textbf{signature}: \text{static/derived} \quad f: \text{Prod}(D_1, D_2, \dots, D_n) \to D_0 \\ \textbf{definitions}: \textbf{function} \ f(d_1 \ \text{in} \ D_1, d_2 \ \text{in} \ D_2, \dots d_n \ \text{in} \ D_n) = R \end{pmatrix}$$
$$\Rightarrow \ \textbf{fun} \ f(\llbracket D_1 \rrbracket \ d_1, \llbracket D_2 \rrbracket \ d_2, \dots, \llbracket D_n \rrbracket \ d_n) \ \textbf{returns} \ \llbracket D_0 \rrbracket = \delta(R)$$

where R is the body of the function. Our approach supports only the conditional and case terms and the boolean terms in these functions. The mapping function for them is defined as follows:

$$\delta \begin{pmatrix} \textbf{if} \ (c) \ \textbf{then} \\ R_1 \\ \textbf{else} \\ R_2 \\ \textbf{endif} \end{pmatrix} \Rightarrow \delta(c) \ ? \ \delta(R_1) : \delta(R_2)$$

$$\delta \begin{pmatrix} \textbf{switch} \ T \\ \quad \textbf{case} \ v_1 : R_1 \\ \quad \dots \\ \quad \textbf{case} \ v_n : R_n \\ \quad \textbf{otherwise} \ R_{ow} \\ \textbf{endswitch} \end{pmatrix} \Rightarrow \begin{matrix} \delta(T) == \delta(v_1) \ ? \ \delta(R_1) : \\ \dots \\ \delta(T) == \delta(v_n) \ ? \ \delta(R_n) : \\ \delta(R_{ow}) \end{matrix}$$

where c is a boolean expression and T is a term.

*Controlled functions:* Against the static and derived functions, controlled functions are not fixed during the execution of the model and may be changed. In fact, transition rules update them. In order to map a controlled function into BIR, we decompose it into its locations and we define a BIR variable for each location. So, for an n-ary controlled function $f{:}Prod(D_1,...,D_n){\to}D_0$, if $D_1$ is a record, then we define $\prod_{2\leq i\leq n}|D_i|$ variables and add them to the record $D_1$, otherwise if $D_j$ is a record or not, when $2{<}j{\leq}n$, we define $\prod_{1\leq i\leq n}|D_i|$ variables. The variables name is $f\_eD_1\_eD_2\_..\_eD_n$ where $eD_1{\in}D_1$, ..., $eD_n{\in}D_n$. Notice that, the product of the cardinality of the domains of a function determines the number of the corresponding variables and the codomain of a function, instead, determines the type of the variables.

*Monitored functions:* These functions are usually used to implement an interaction between the model and the user. Their values can be passed to it by an environment file or by command line. Since the interaction between the environment and Bogor is not possible, hence we define a variable for each monitored function as described for nullary and controlled functions, and create the inactive threads which are used to generate the state space of the monitored functions, we call them *monitored threads*. They are being active in the first location of main thread, loc0.

For example, the monitored function $mon{:}[(D)^*{\to}]D_0$ must be decomposed into its locations; for each location, we create a monitored thread. The cardinality of the domain of the monitored function determines the number of the corresponding monitored threads in BIR. The codomain of the function, instead, determines the type of the monitored locations. Therefore, the translation rule for monitored function, *mon*, is defined as follows:

$$\delta(\mathbf{monitored}\ mon{:}[(D)^*{\to}]D_0)\ \Rightarrow\ \begin{array}{l}\mathbf{thread}\ mon_j\_monitored()\ \{\ \mathbf{loc}\ loc0:\\ \qquad\mathbf{do}\ \{mon_j := v_1;\}\mathbf{goto}\ loc0;\\ \qquad\cdots\\ \qquad\mathbf{do}\ \{mon_j := v_n;\}\ \mathbf{goto}\ loc0;\ \}\end{array}$$

where $v_i \in D_0$, $1 \leq i \leq |D_0|$, and $mon_j \in mLocs$, and *mLocs* is the set of the monitored locations.

```
thread mon_monitored() {
        loc loc0 :
                do {mon:=true;}goto loc0;
                do {mon:=false;}goto loc0;
        }
```

**Fig. 9.** BIR monitored functions.

When Bogor encounters an active monitored thread it creates a state for each value of the variable. Bogor sets the values of monitored variables at the beginning of the transaction. It causes transition rules deal with the monitored location values of the current state. Fig. 9 shows an example for the boolean variable *mon* in BIR notation.

## 6.3. Initialization

In the current step of our encoding, we initialize the data structure of the ASM model which consists of static, derived, monitored and controlled functions. Since the value of the static functions and the computation mechanism of the derived functions are always fixed and the monitored functions can take all values through monitored threads, hence we only initialize the controlled functions in this section. Although the initialization of all controlled functions are optional, however, missing one of them may cause an unwanted situation because, when Bogor encounters an empty variable, it chooses the first value of its domain as its initial value which may be incorrect and undesirable. So, we consider it completely. Initialization of the controlled functions is expressed in the *default init* section in ASM model. Our approach supports only simple initialization, group initialization, and conditional initialization. We translate them into BIR and put them in loc0, i.e., the initial location. The translation rules for them are defined as follows:

$$\delta(\textbf{function } T = v) \Rightarrow \langle\!\langle T \rangle\!\rangle = \delta(v)$$

$$\delta(\textbf{function } T(\$c \text{ } in \text{ } D) = v) \Rightarrow \langle\!\langle T(c) \rangle\!\rangle = \delta(v) , \forall c \in D$$

$$\delta \begin{pmatrix} \textbf{function } T(\$c \text{ } in \text{ } D) = \textbf{switch} \text{ } (\$c) \\ \textbf{case } c_1 : v_1 \\ ... \\ \textbf{case } c_n : v_n \\ \textbf{endswitch} \end{pmatrix} \Rightarrow \langle\!\langle T(c) \rangle\!\rangle = \delta(v) , \forall c \in \{c_1, ..., c_n\} \subseteq D$$

## 6.4. Rules

After encoding the data structures and initializing them, we encode the rules. The behaviour of the ASM model is expressed by rules, which describe how the controlled functions change their interpretation from one state to another state. Since the main rule is the starting point of the execution, so we start from it and continue executing a depth visit of all the rules we encounter.

We first create a new location in the main thread, *loc1*, which is used to translate the main rule into BIR, and we jump from *loc0* to this location. We also use a global stack, we call it *Guards*, which stores all conditions that we encounter (e.g., conditional rule, case rule, etc). If we want to briefly describe our transition rule encoding, we can say: we starts our encoding in the main rule, (a) we push the boolean conditions on the stack *Guards*, (b) we create a suitable BIR guarded command when we encounter an update rule, (c) and finally, we remove the condition when we leave its scope. The BIR commands are stated in guard command format (Fig. 10): when *guard expression* is satisfied, *actions* are executed. In Bogor, all guards in an active location are evaluated simultaneously: if more than one guard in the active location is true, then Bogor chooses non-deterministically one of them. We define different locations with the keyword *loc* and explicitly jump among them with *goto*.

```
loc loc_m:
       when (guarded_expression) do [invisible] {actions;} goto loc_n;
```

**Fig. 10.** BIR guarded command.

The details of our encoding procedure to transform the transition rules which have been introduced in Section 4 are as follows:

*Update rule:* Update rule is the simplest rule. In our encoding approach, when we encounter the update rule, we create a BIR guarded command. The guard expression and action are made by conjunction of the contents of the stack *Guards* and translation of update rule, respectively. The translation rule for the update rule is defined as follows:

$$\delta(f(t_1, \ldots, t_n) := t_0) \Rightarrow \textbf{when}(\bigwedge_{i=1}^{|SC|} c[i]) \textbf{ do [invisible] } \{ \langle\!\langle f(v_1, \ldots, v_n) \rangle\!\rangle := \delta(v_0); \} \textbf{ goto } loc;$$

where *SC* is the stack counter, and *c[i]* is the $i^{th}$ stack element, and *loc* is the current BIR location. Moreover, *f* is a function symbol and $t_0, t_1, \ldots, t_n$ are terms which are evaluated in the current state of the model yielding to values $v_0, v_1, \ldots, v_n$.

*Conditional rule:* To encode the conditional rules, we push $\delta(cond)$ on the stack and visit $R_{if}$ (i.e., $\delta(R_{if})$). If else branch is not null, we remove $\delta(cond)$ from the stack and we push its negation (i.e., $\delta(\neg cond)$) on the stack. Then, we visit $R_{else}$ (i.e., $\delta(R_{else})$). Finally, when we encounter *endif*, we remove the last condition (i.e., $\delta(cond)$ or $\delta(\neg cond)$) from the stack.

*Case rule:* To encode the case rules:
- For each case branch *i*:
  - We push $(\wedge_{1 \leq j < i} \delta(T)!= \delta(v_j)) \wedge (\delta(T)== \delta(v_i))$ on the stack.
  - $R_i$ is visited (i.e., $\delta(R_i)$).
  - We remove the previous condition from the stack.
- If otherwise branch is not null
  - We push $\wedge_{1 \leq i < n} \delta(T)!= \delta(v_i)$ on the stack.
  - $R_{ow}$ is visited (i.e., $\delta(R_{ow})$).
  - We remove the previous condition from the stack.

*Choose rule:* Choose rule is used to model arbitrary selection of processes at a high level of abstraction. It executes $R_{1,\ldots,n}$ one time for some $V_k$ which satisfies $cond_{1,\ldots,n}$ where $V_k$ is the $k^{th}$ element of the set *Values*={$(v_1,\ldots,v_n) \mid v_1{\in}D_1,\ldots,v_n{\in}D_n$}. Suppose that the number of branches in the choose rule is $nB=\prod_{1 \leq i \leq n}|D_i|$, where $|D_i|$ is the cardinality of $i^{th}$ domain. For each branch ($\forall k \in 1..nB$):

- We replace the variables $d_1, \ldots, d_n$ of $cond_{1,\ldots,n}$ and $R_{1,\ldots,n}$ with the current values $V_k$, and called them $cond_k$ and $R_k$, respectively.
- We push $\delta(cond_k)$ on the stack *Guards*.
- $R_k$ is visited (i.e., $\delta(R_k)$).
- We remove $\delta(cond_k)$ from the stack.

If *ifnone* branch is present, we push $\wedge_{1 \leq k \leq nB}!\,\delta(cond_k)$ on the stack Guards, visit the $R_{ifnone}$ (i.e., $\delta(R_{ifnone})$), and remove the last condition from the stack.

*Parallel rules:* AsmetaL supports of two parallel rules which allow for rules to be grouped together and executed in parallel, the par block rule and the forall rule.

*Par block rule:* For the par block rule **par** $R_1$ ... $R_n$ **endpar**, we define $\sum_{i=2}^{n} 2^{i-1}$ consecutive locations. We also use a set, we call it *changed*, which stores all variables that we update. When we want to update a variable, we check the existence of it in the set *changed*. For example, assume that $R_1$ updates the variable $v_1$ in the current state (i.e., *changed=$\{v_1\}$*), thus $R_2, ... , R_n$ could not change the value of the variable $v_1$ in the this state. Hence, for the transition rule $R_1$, we create two invisible BIR guarded command. In the first BIR guarded command with the guard $c$, $R_1$ is executed and we jump to the location $L_{pid,1}$, where *pid* identifies the different par block rules. We also add the variables which are updated by $R_1$ to the set *changed*. In the second BIR guarded command with the guard $\neg c$, we only jump to the location $L_{pid,2}$. The transition rule $R_2$ is visited in the locations $L_{pid,1}$ and $L_{pid,2}$, if no variable exists in *changed* set that is updated by $R_2$.

In general, in the current location, we visit the transition rule $R_1$ and jump to $L_{pid,1}$ or $L_{pid,2}$. In the $i^{th}$ location, we visit the transition rule $R_i$ invisibly if it updates the variables which do not exist in the set *changed*, and jump to the location $L_{pid,j}$, where $2^{k-2} \leq i \leq 2^{k-1}$ and $2^{k-1} \leq j \leq 2^k$ and $2 \leq k < n$. Due to the use of invisible BIR guarded commands in these locations, Bogor generates no state. In the last locations (i.e., $L_{pid,2^{n-2}}, ... , L_{pid,2^{n-1}}$), we visit $R_n$ invisibly if it updates the variables which are not exist in the set *changed* and jump to the next location (i.e., the first location of the next rule). We define an empty visible BIR guarded command at the end of the model, we call it *endloc*. When Bogor encounters the visible BIR guarded command, it generates a state for representing it. The translation rule for the par block rule is defined as follows:

$$\delta(\textbf{par } R_1 ... R_n \textbf{ endpar}) \Rightarrow \delta_{cLoc}^{invisible}(R_1),$$

$$\delta_{L_{pid,1},L_{pid,2}}^{invisible}(R_2 | \forall v_i \in R_2 \rightarrow v_i \notin changed),$$

$$...,$$

$$\delta_{L_{pid,2^{n-2}},...,L_{pid,2^{n-1}}}^{invisible}(R_n | \forall v_i \in R_n \rightarrow v_i \notin changed)$$

For example, Part *a* of Fig. 11 shows the par block rule which are consisted of two conditional rules $R_1$ and $R_2$. To encode the par block rule, we define $\sum_{i=2}^{2} 2^{i-1} = 2$ locations which are labeled by *loc1_1* and *loc1_2*. For the first conditional rule, we create two invisible BIR guarded command in the current location (i.e., loc1): the first BIR guarded command checks whether the variable *foo* is equal to 10 (*foo==10*), and, if it is the case, sets its base value (*foo:=1*), and jumps to *loc1_1*, whereas the second BIR guarded command checks whether *foo* is not equal to 10, only jumps to *loc1_2* and does nothing. Since the conditional rule $R_2$ changes the variable *foo*, we must visit it in the location which its changed set does not have the variable *foo* (i.e., *loc1_2*). In the location *loc1_1*, we define an empty invisible BIR guarded command and jump to endloc.

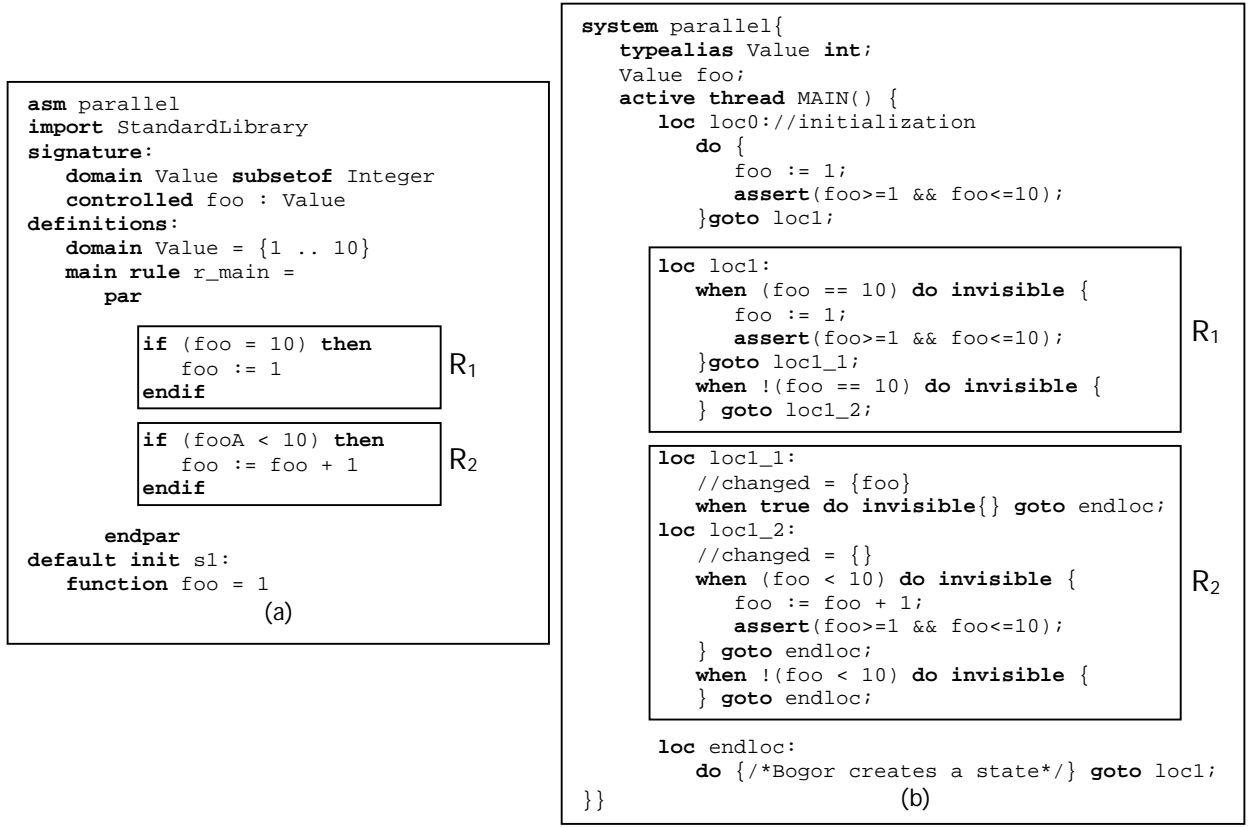

**Fig. 11.** Par block rule, (a) AsmetaL notation, (b) BIR notation.

*Forall rule:* The forall rules allow the same rule to be executed in parallel for all values of a domain that satisfy a condition: hence they are considered as special par block rules. There is only a difference in the mapping of forall rules and par block rules: for each branch $k$, we push $cond_k$ on the stack *Guards*, visit $R_k$, and remove $cond_k$ from the stack, where $k \in 1..\prod_{1 \le i \le n} |D_i|$. Notice that, definition of $V_k$, $R_k$, and $cond_k$ are the same with the choose rule. So, our encoding for the forall rule is as following:

$$\delta(\textbf{forall } d_1 \textbf{ in } D_1, \dots, d_n \textbf{ in } D_n \textbf{ [with } \langle cond[d_1, \dots, d_n] \rangle] \textbf{ do } R[d_1, \dots, d_n])$$
$$\Rightarrow \delta(\textbf{par } R_{1|cond_1} \dots R_{n|cond_n} \textbf{ endpar})$$

*Sequential block rule:* Bogor sequentially evaluates the consecutive locations which are connected by the goto command. Hence, to encode the seq block rule, we define $n$ consecutive locations which are labelled by $L_{seqid,i}$. Before the seq rule is encoded, we create an invisible BIR guarded command in which the guard expression is made by conjunction of the stack Guards, and the action is empty; then we jump to the first location of the seq rule. In the $i^{th}$ location, we visit the transition rule $R_i$ and jump to the location i+1, where $1 \le i < n$. In the last location, we visit $R_n$ and jump to the first location of the next rule or to the *endloc*. Notice that the contents of the stack *Guards* are the same for the block rules (i.e., par block and seq block). The translation rule for the seq block rule is defined as follows:

$$\delta(\textbf{seq} \quad R_1 \quad \dots \quad R_n \ \textbf{endseq}) \ \Rightarrow \ \delta_{L_{sid.1}}(R_1), \delta_{L_{sid.2}}(R_2), \dots, \delta_{L_{sid.n}}(R_n)$$

where sid identifies the different seq block rules.

*Let rule:* The transition rule for the let rules is:

$$\delta(\textbf{let} \ (\$x_i = V_i)^+ \ R \ \textbf{endlet}) \ \Rightarrow \ \delta(R_{V_i})$$

where $R_{Vi}$ is the rule where the variables $\$x_i$ has been replaced with the values $V_i$. So, to encode the let rule, we must visit the rule $R_{Vi}$.

*Skip rule:* The skip rule is a no-operation rule in which no location is changed. Since, BIR also uses "*skip*" as the empty rule, we do not define any further command to consider it.

*Expressions:* ASM2Bogor supports the expression types supported in the AsmetaL language. The expression is made of the function terms and operators. The function terms must be translated into the BIR variables. Table 3 shows the supported operator.

**Table 3.** AsmetaL to BIR logical operator conversion.

| AsmetaL operator | BIR operator |
|:---:|:---:|
| + | + |
| - | - |
| * | * |
| / | / |
| mod | % |
| = | == |
| != | != |
| < | < |
| <= | <= |
| > | > |
| >= | >= |
| and | && |
| or | \|\| |
| not | ! |

We also support forall and exists terms. To encode the forall term, we extend the condition of the term into a conjunction of the elements of the forall domains. We assume that the definition of $k$ and $cond_k$ is the same with the choose rule; so, the translation rule for the expression is:

$$\delta(\textbf{forall} \ d_1 \ \textbf{in} \ D_1, \dots, d_n \ \textbf{in} \ D_n \ [\textbf{with} \ \langle cond[d_1, \dots, d_n] \rangle]) \ \Rightarrow \ \bigwedge_{i=1}^{k} \delta(cond_i)$$

Similarly, to encoding the exists expression, we extend the condition of the expression into a disjunction of the elements of the exists domains.

$$\delta(\textbf{exists} \ d_1 \ \textbf{in} \ D_1, \dots, d_n \ \textbf{in} \ D_n \ [\textbf{with} \ \langle cond[d_1, \dots, d_n] \rangle]) \ \Rightarrow \ \bigvee_{i=1}^{k} \delta(cond_i)$$

### 6.5.   Linear temporal logic properties

In this section we describe the declaration procedure for Linear Temporal Logic (LTL) properties with AsmetaL. In AsmetaL, the library *LTLlibrary* contains functions that represent the LTL temporal operators. The library must be imported when we want to use LTL operators in the AsmetaL model. Table 4 shows all the supported LTL operators.

**Table 4.** AsmetaL to BIR LTL operator conversion.

| AsmetaL function | BIR LTL operator | comment |
|---|---|---|
| **g**(p) | always(p) | In each state of the model the proposition *p* must be satisfied. |
| **f**(p) | eventually(p) | Eventually, there is a state in which the proposition *p* is satisfied. |
| p **u** q | until(p, q) | *p* must hold at least until *q* holds |
| p **v** q | release(p, q) | *q* must remain true until and including the point where *p* becomes true. |

```
asm checkAxiomAndProperty
import StandardLibrary
import LTLlibrary

signature:
    dynamic controlled m:Boolean
    dynamic controlled n:Boolean

definitions:

    LTLSPEC NAME ltl_neverEQ:=not(f(m=n))

    //invariant over m,n: m!=n
    LTLSPEC NAME ltl_inv:=g(m!=n)

    main rule r_Main =
        par
            m:=not(n)
            n:=not(m)
        endpar

default init s0:
    function m = true
    function n = false                    (a)
```

```
system checkAxiomAndProperty {

fun ltl_neverEQ() returns boolean =
    LTL.temporalProperty(
        Property.createObservableDictionary(
            Property.createObservableKey("P", m==n)),
        LTL.negation(LTL.eventually(LTL.prop("P")))
    );

fun ltl_inv() returns boolean =
    LTL.temporalProperty(
        Property.createObservableDictionary(
            Property.createObservableKey("P", m!=n)),
        LTL.always(LTL.prop("P"))
    );
boolean m;//controlled
boolean n;//controlled
main thread MAIN() {
    loc loc0://initialization
        do {
            m := true;
            n := false;
        } goto loc1;
    loc loc1:
        do invisible {
            m := !(n);
            n := !(m);
        } goto endloc;
    loc endloc:
        do {/*Bogor creates a state*/} goto loc1;
    }
}                                         (b)
```

**Fig. 12.** LTL properties and invariants encoding: (a) AsmetaL code, (b) BIR code

The user can check safety and liveness properties by defining the desired property through the *LTLSPEC NAME ltl_SpecName:= P* format in an AsmetaL model, where *P* is an LTL formula. These properties must be declared before the main rule. Since, Bogor can check deadlock directly, we do not consider any further command to consider it.

We also support AsmetaL invariants. An AsmetaL invariant is a property that must be verified in each state. In order to check an invariant in Bogor, the invariant

$$\text{invariant over } \text{id}_1, \dots, \text{id}_n : P_{id_1, \dots, id_n}$$

must be rewritten as *LTLSPEC NAME Itl_invName:= g(P$_{id1, \dots, idn}$)* format. *P$_{id1, \dots, idn}$* is a boolean expression which expresses the constraint over domains, functions or rules *id$_1$, ... ,id$_n$*.

We translate the LTL specification to the boolean BIR function which is then checked by Bogor. For example, Part *b* of Fig. 12 shows the two equivalent BIR functions of the *Itl_neverEQ* and *Itl_inv* specifications which are represented in part *a* of Fig. 12. *Itl_neverEQ* is a safety property which is true if there is no state in which the controlled location *m* is equal to *n* and *Itl_inv* is an AsmetaL invariant that is equivalent to *Itl_neverEQ*.

Let's check the deadlock and correctness of the desired properties:

```
user@linux-cdhm:~> asm2bogor.sh checkAxiomAndProperty.asm
Bogor v.1.2 (build <version>)
(c) Copyright by Kansas State University

Web: http://bogor.projects.cis.ksu.edu

Transitions: 3, States: 2, Matched States: 1, Max Depth: 3, Errors found: 0, Used Memory: 0MB
** checkAxiomAndProperty is not in DEADLOCK
**LTLSPEC NAME Itl_neverEQ:= not(f(m=n)) is true
**LTLSPEC NAME Itl_inv:= g(m!=n) is true
Done!
```

## 7. Case Studies

In this section, we present the results of applying ASm2Bogor on several case studies. After a short introduction to the case studies, we describe the ASM models written in AsmetaL language and do not have the unsupported elements. Finally, we encode the models to BIR along with the desired properties to be verified.

### 7.1. Sluice gate control

Sluice gate control problem is a simple irrigation system, introduced in [3]. It uses a small sluice, with a rising and a falling gate. A computer system controls the sluice gate. The description of the requirements is that:

> "The gate should be held in the fully open position for ten minutes in every three hours and otherwise kept in the fully closed position. The gate is opened and closed by rotating vertical screws. The screws are driven by a small *motor*, which can be controlled by clockwise, anticlockwise, on and off pulses.
>
> There are *sensors* at the top and bottom of the gate travel; at the top it's fully open, at the bottom it's fully shut.
>
> The *connection to the computer* consists of four pulse lines for motor control and two status lines for the gate sensors." [3]

The system is in a fully closed, opening, fully opened, or closing cycle as shown in Fig. 13: Whenever the gate is fully closed and the time 170(min) has passed, the computer system sets the screws direction to clockwise and turns on the motor, and the gate switches to the opening phase. When the gate has reached its top position, the computer system turns off the motor and the gate enters phase fully opened. After ten minutes, the computer system sets the screws direction to anticlockwise and turns on the motor again, and the gate switches to the closing phase. Finally, the computer system turns off the motor and the gate back to the fully closed phase whenever the bottom sensor shows the gate has reached its lower position. At the beginning the gate is fully closed and the motor is off.

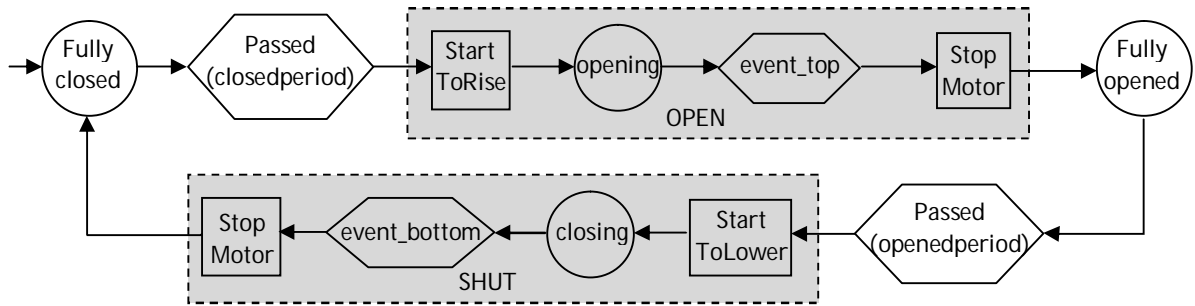

**Fig. 13.** Sluice gate control

Fig. 14 shows the AsmetaL model of the sluice gate control problem.

```
asm sluiceGateControl

import StandardLibrary
import LTLlibrary

signature:
    domain Minutes subsetof Integer
    enum domain PhaseDomain = { FULLYCLOSED | OPENING | FULLYOPENED | CLOSING }
    enum domain DirectionDomain = { CLOCKWISE | ANTICLOCKWISE }
    enum domain MotorDomain = { ON | OFF }
    dynamic controlled phase: PhaseDomain
    dynamic controlled dir: DirectionDomain
    dynamic controlled motor: MotorDomain
    dynamic monitored passed: Minutes -> Boolean
    dynamic monitored event_top: Boolean
    dynamic monitored event_bottom: Boolean

definitions:
    domain Minutes = {10, 170}

    rule r_start_to_raise =
        par
            dir := CLOCKWISE
            motor := ON
        endpar

    rule r_start_to_lower =
        par
            dir := ANTICLOCKWISE
            motor := ON
        endpar

    rule r_stop_motor =
        motor := OFF

//liveness: correctness of the transitions between states
LTLSPEC NAME ltl_opened2closing:=g((phase=FULLYOPENED and passed(10)) implies f(phase=CLOSING))
LTLSPEC NAME ltl_closing2closed:=g((phase=CLOSING and event_bottom) implies f(phase=FULLYCLOSED))
LTLSPEC NAME ltl_closed2opening:=g((phase=FULLYCLOSED and passed(170)) implies f(phase=OPENING))
LTLSPEC NAME ltl_opening2openend:=g((phase=OPENING and event_top) implies f(phase=FULLYOPENED))
```

```
//properties on the relationship between the state and the motor
LTLSPEC NAME ltl_motorOFF1:=g(phase=FULLYCLOSED  implies  motor=OFF)
LTLSPEC NAME ltl_motorOFF2:=g(phase=FULLYOPENED  implies  motor=OFF)
LTLSPEC NAME ltl_motorON1:=g(phase=OPENING implies (motor=ON and dir=CLOCKWISE))
LTLSPEC NAME ltl_motorON2:=g(phase=CLOSING implies (motor=ON and dir=ANTICLOCKWISE))

    main rule r_Main =
        par
            if(phase=FULLYCLOSED) then
                if(passed(170)) then
                    par
                        r_start_to_raise[]
                        phase := OPENING
                    endpar
                endif
            endif
            if(phase=OPENING) then
                if(event_top) then
                    par
                        r_stop_motor[]
                        phase := FULLYOPENED
                    endpar
                endif
            endif
            if(phase=FULLYOPENED) then
                if(passed(10)) then
                    par
                        r_start_to_lower[]
                        phase := CLOSING
                    endpar
                endif
            endif
            if(phase=CLOSING) then
                if(event_bottom) then
                    par
                        r_stop_motor[]
                        phase := FULLYCLOSED
                    endpar
                endif
            endif
        endpar

default init s0:
   function phase = FULLYCLOSED
   function motor = OFF
```

**Fig. 14.** Sluice gate control: AsmetaL model

We define a controlled function *phase* for recording the state of the gate: the gate can be fully closed, opening, fully opened, or closing. We also introduce two controlled locations, *motor* and *dir*, that model the motor is turned on or off, and the direction of the screws, clockwise or anticlockwise.

The boolean monitored locations *Passed(10)* and *Passed(170)* indicate when respectively the fully opened period and the fully closed period have passed. Moreover, the boolean monitored locations *event_top* and *event_bottom* say if the gate has reached its top and bottom position.

In the model we have defined some properties.

To check the correctness of the transitions between states we declare the four liveness properties:

*LTLSPEC NAME ltl_opened2closing:=g((phase=FULLYOPENED and passed(10)) implies f(phase=CLOSING))*
*LTLSPEC NAME ltl_closing2closed:=g((phase=CLOSING and event_bottom) implies f(phase=FULLYCLOSED))*
*LTLSPEC NAME ltl_closed2opening:=g((phase=FULLYCLOSED and passed(170)) implies f(phase=OPENING))*

*LTLSPEC NAME ltl_opening2openend:=g((phase=OPENING and event_top)implies f(phase=FULLYOPENED))*

For example, if the gate is fully opened (*phase=FULLYOPENED*) and the opened period has passed (*passed(10)*), it becomes opening (*phase=OPENING*).

The safety properties

*LTLSPEC NAME ltl_motorOFF1:= g(phase=FULLYCLOSED implies motor=OFF)*
*LTLSPEC NAME ltl_motorOFF2:= g(phase=FULLYOPENED implies motor=OFF)*

check that, if the gate is stopped then the motor must be turn off.

The two safety properties

*LTLSPEC NAME ltl_motorON1:= g(phase=OPENING implies (motor=ON and dir=CLOCKWISE))*
*LTLSPEC NAME ltl_motorON2:= g(phase=CLOSING implies (motor=ON and dir=ANTICLOCKWISE))*

check that, if the gate is moving then the motor is turned on, and the screws direction is clockwise when the gate is opening, and is anticlockwise when the gate is closing.

Finally, we want to check the absence of deadlock. Since, Bogor can check deadlock directly, we do not consider any further command to consider it.

Fig 15 is the BIR code obtained from the mapping of the AsmetaL code represented in Fig 14.

```
system sluiceGateControl {

    fun ltl_opened2closing() returns boolean =
        LTL.temporalProperty(
            Property.createObservableDictionary(
                Property.createObservableKey("P1", phase==PhaseDomain.FULLYOPENED),
                Property.createObservableKey("P2", passed_10),
                Property.createObservableKey("P3", phase==PhaseDomain.CLOSING)),
            LTL.always(
                LTL.implication(
                    LTL.conjunction(LTL.prop("P1"),LTL.prop("P2")),
                    LTL.eventually(LTL.prop("P3"))
                )
            )
        );
    fun ltl_closing2closed() returns boolean =
        LTL.temporalProperty(
            Property.createObservableDictionary(
                Property.createObservableKey("P1", phase==PhaseDomain.CLOSING),
                Property.createObservableKey("P2", event_bottom),
                Property.createObservableKey("P3", phase==PhaseDomain.FULLYCLOSED)),
            LTL.always(
                LTL.implication(
                    LTL.conjunction(LTL.prop("P1"),LTL.prop("P2")),
                    LTL.eventually(LTL.prop("P3"))
                )
            )
        );
    fun ltl_closed2opening() returns boolean =
        LTL.temporalProperty(
            Property.createObservableDictionary(
                Property.createObservableKey("P1", phase==PhaseDomain.FULLYCLOSED),
                Property.createObservableKey("P2", passed_170),
                Property.createObservableKey("P3", phase==PhaseDomain.OPENING)),
            LTL.always(
                LTL.implication(
                    LTL.conjunction(LTL.prop("P1"),LTL.prop("P2")),
                    LTL.eventually(LTL.prop("P3"))
                )
            )
        );
    fun ltl_opening2openend() returns boolean =
        LTL.temporalProperty(
```

```
        Property.createObservableDictionary(
            Property.createObservableKey("P1", phase==PhaseDomain.OPENING),
            Property.createObservableKey("P2", event_top),
            Property.createObservableKey("P3", phase==PhaseDomain.FULLYOPENED)),
        LTL.always(
            LTL.implication(
                LTL.conjunction(LTL.prop("P1"),LTL.prop("P2")),
                LTL.eventually(LTL.prop("P3"))
            )
        )
    );
fun ltl_motorOFF1() returns boolean =
    LTL.temporalProperty(
        Property.createObservableDictionary(
            Property.createObservableKey("P1", phase==PhaseDomain.FULLYCLOSED),
            Property.createObservableKey("P2", motor==MotorDomain.OFF)),
        LTL.always(LTL.implication(LTL.prop("P1"),LTL.prop("P2")))
    );
fun ltl_motorOFF2() returns boolean =
    LTL.temporalProperty(
        Property.createObservableDictionary(
            Property.createObservableKey("P1", phase==PhaseDomain.FULLYOPENED),
            Property.createObservableKey("P2", motor==MotorDomain.OFF)),
        LTL.always(LTL.implication(LTL.prop("P1"),LTL.prop("P2")))
    );
fun ltl_motorON1() returns boolean =
    LTL.temporalProperty(
        Property.createObservableDictionary(
            Property.createObservableKey("P1", phase==PhaseDomain.OPENING),
            Property.createObservableKey("P2", motor==MotorDomain.ON),
            Property.createObservableKey("P2", dir==DirectionDomain.CLOCKWISE)),
        LTL.always(
            LTL.implication(LTL.prop("P1"),LTL.conjunction(LTL.prop("P2"),LTL.prop("P3")))
        )
    );
fun ltl_motorON2() returns boolean =
    LTL.temporalProperty(
        Property.createObservableDictionary(
            Property.createObservableKey("P1", phase==PhaseDomain.CLOSING),
            Property.createObservableKey("P2", motor==MotorDomain.ON),
            Property.createObservableKey("P2", dir==DirectionDomain.ANTICLOCKWISE)),
        LTL.always(
            LTL.implication(LTL.prop("P1"),LTL.conjunction(LTL.prop("P2"),LTL.prop("P3")))
        )
    );

typealias Minutes int;
enum PhaseDomain {FULLYCLOSED, OPENING, FULLYOPENED, CLOSING}
enum DirectionDomain {CLOCKWISE, ANTICLOCKWISE}
enum MotorDomain {ON, OFF}
PhaseDomain phase;//controlled
DirectionDomain dir;//controlled
MotorDomain motor;//controlled
boolean passed_10;//monitored
boolean passed_170;//monitored
boolean event_top;//monitored
boolean event_bottom;//monitored

thread passed_10_monitored(){
    loc loc0:
        do {passed_10:=true;} goto loc0;
        do {passed_10:=false;} goto loc0;
}
thread passed_170_monitored(){
    loc loc0:
        do {passed_170:=true;} goto loc0;
        do {passed_170:=false;} goto loc0;
}
thread event_top_monitored(){
    loc loc0:
        do {event_top:=true;} goto loc0;
```

```
        do {event_top:=false;} goto loc0;
}
thread event_bottom_monitored(){
    loc loc0:
        do {event_bottom:=true;} goto loc0;
        do {event_bottom:=false;} goto loc0;
}

active thread MAIN(){
    loc loc0://initialization
        do {
            phase:=PhaseDomain.FULLYCLOSED;
            motor:=MotorDomain.OFF;

            //monitored functions
            start passed_10_monitored();
            start passed_170_monitored();
            start event_top_monitored();
            start event_bottom_monitored();
        }goto loc1;

    //rules
    //FULLYCLOSED -> OPENING
    loc loc1:
        when (phase==PhaseDomain.FULLYCLOSED && passed_170) do invisible {
            dir := DirectionDomain.CLOCKWISE;
            motor := MotorDomain.ON;
            phase := PhaseDomain.OPENING;
        }goto loc2_C1;
        when !(phase==PhaseDomain.FULLYCLOSED && passed_170) do invisible {} goto loc2_N1;

    //OPENING -> FULLYOPENED
    loc loc2_C1:
        //changed={dir, motor, phase}
        do invisible{} goto loc2_C1_N2;
    loc loc2_N1:
        //changed={}
        when (phase==PhaseDomain.OPENING && event_top) do invisible {
            motor := MotorDomain.OFF;
            phase := PhaseDomain.FULLYOPENED;
        }goto loc2_N1_C2;
        when !(phase==PhaseDomain.OPENING && event_top) do invisible {} goto loc2_N1_N2;

    //FULLYOPENED -> CLOSING
    loc loc2_C1_C2:
        //changed={dir, motor, phase}
        do invisible{} goto loc2_C1_C2_N3;
    loc loc2_C1_N2:
        //changed={dir, motor, phase}
        do invisible{} goto loc2_C1_N2_N3;
    loc loc2_N1_C2:
        //changed={motor, phase}
        do invisible{} goto loc2_N1_C2_N3;
    loc loc2_N1_N2:
        //changed={}
        when (phase==PhaseDomain.FULLYOPENED && passed_10) do invisible {
            dir := DirectionDomain.ANTICLOCKWISE;
            motor := MotorDomain.ON;
            phase := PhaseDomain.CLOSING;
        }goto loc2_N1_N2_C3;
        when !(phase==PhaseDomain.FULLYOPENED && passed_10) do invisible {} goto loc2_N1_N2_N3;

    //CLOSING -> FULLYCLOSED
    loc loc2_C1_C2_C3:
        //changed={dir, motor, phase}
        do invisible {} goto endloc;
    loc loc2_C1_C2_N3:
        //changed={dir, motor, phase}
        do invisible {} goto endloc;
    loc loc2_C1_N2_C3:
        //changed={dir, motor, phase}
```

```
        do invisible {} goto endloc;
    loc loc2_C1_N2_N3:
        //changed={dir, motor, phase}
        do invisible {} goto endloc;
    loc loc2_N1_C2_C3:
        //changed={motor, phase}
        do invisible {} goto endloc;
    loc loc2_N1_C2_N3:
        //changed={motor, phase}
        do invisible {} goto endloc;
    loc loc2_N1_N2_C3:
        //changed={dir, motor, phase}
        do invisible {} goto endloc;
    loc loc2_N1_N2_N3:
        //changed={}
        when (phase==PhaseDomain.CLOSING && event_bottom) do invisible {
            motor := MotorDomain.OFF;
            phase := PhaseDomain.FULLYCLOSED;
        }goto endloc;
        when !(phase==PhaseDomain.CLOSING && event_bottom) do invisible {} goto endloc;
    loc endloc:
        do {/*Bogor creates a state*/} goto loc1;
    }
}
```

**Fig. 15.** Sluice gate control: BIR model

Let's see the execution of BIR code in order to check the deadlock and correctness of the desired properties:

user@linux-cdhm:~> asm2bogor.sh sluiceGateControl.asm
Bogor v.1.2 (build <version>)
(c) Copyright by Kansas State University

Web: http://bogor.projects.cis.ksu.edu

Transitions: 1041, States: 81, Matched States: 641, Max Depth: 132, Errors found: 0, Used Memory: 0MB
** sluiceGateControl is not in DEADLOCK
**LTLSPEC NAME ltl_opened2closing:=g((phase=FULLYOPENED and passed(10)) implies f(phase=CLOSING)) is true
**LTLSPEC NAME ltl_closing2closed:=g((phase=CLOSING and event_bottom) implies f(phase=FULLYCLOSED)) is true
**LTLSPEC NAME ltl_closed2opening:=g((phase=FULLYCLOSED and passed(170)) implies f(phase=OPENING)) is true
**LTLSPEC NAME ltl_opening2opened:=g((phase=OPENING and event_top) implies f(phase=FULLYOPENED)) is true
**LTLSPEC NAME ltl_motorOFF1:=g(phase=FULLYCLOSED implies motor=OFF) is true
**LTLSPEC NAME ltl_motorOFF2:=g(phase=FULLYOPENED implies motor=OFF) is true
**LTLSPEC NAME ltl_motorON1:=g(phase=OPENING implies (motor=ON and dir=CLOCKWISE)) is true
**LTLSPEC NAME ltl_motorON2:=g(phase=CLOSING implies (motor=ON and dir=ANTICLOCKWISE)) is true
Done!

## 7.2. One way traffic light control

As another case study for model checking ASM by using ASM2Bogor, we chose one way traffic light control [3]. It is a system consists of two traffic light units placed at each end of an alternated one-way street and a computer system that controls them. Each light has two different coloured lights, a red light (Stop light) and a green light (Go light). The cars can pass through the street when the Go light is on, and must be waited when the Stop light is on. The Stop light and the Go light of a traffic light unit are complement. The computer system turns on or off the Stop and Go light by sending two pulses, respectively *RPulse* and *GPulse*. The pulses complement the corresponding lights.

At the beginning both traffic lights show stop. We call the state STOP1STOP2, i.e. both Stop lights are on and both Go lights are off.

As shown in Fig. 16, the traffic lights follow a eight phase cycle:

- for 50 seconds, the system is in STOP1STOP2 phase, then it enters phase STOP1STOP2CHANGING and the computer system turns on the Go light and turns off the Stop light of the second traffic light by sending the R and G pulses to it;
- whenever the time assumed for the changing of the Stop and the Go lights is passed, in the example 10 seconds, the system enters phase GO2STOP1;
- after 120 seconds, the system enters phase GO2STOP1CHANGING and the computer system turns off the Go light and turns on the Stop light of the second traffic light by sending the R and G pulses to it;
- after 10 seconds, the system enters phase STOP2STOP1 (i.e. both traffic light units show stop again);
- after 50 seconds, the system enters phase STOP2STOP1CHANGING and the computer system turns on the Go light and turns off the Stop light of the first traffic light by sending the R and G pulses to it;
- after 10 seconds, the system enters phase GO1STOP2;
- after 120 seconds, the system enters phase GO1STOP2CHANGING and the computer system turns off the Go light and turns on the Stop light of the first traffic light by sending the R and G pulses to it;
- finally, after 10 seconds, the system back to phase STOP1STOP2;

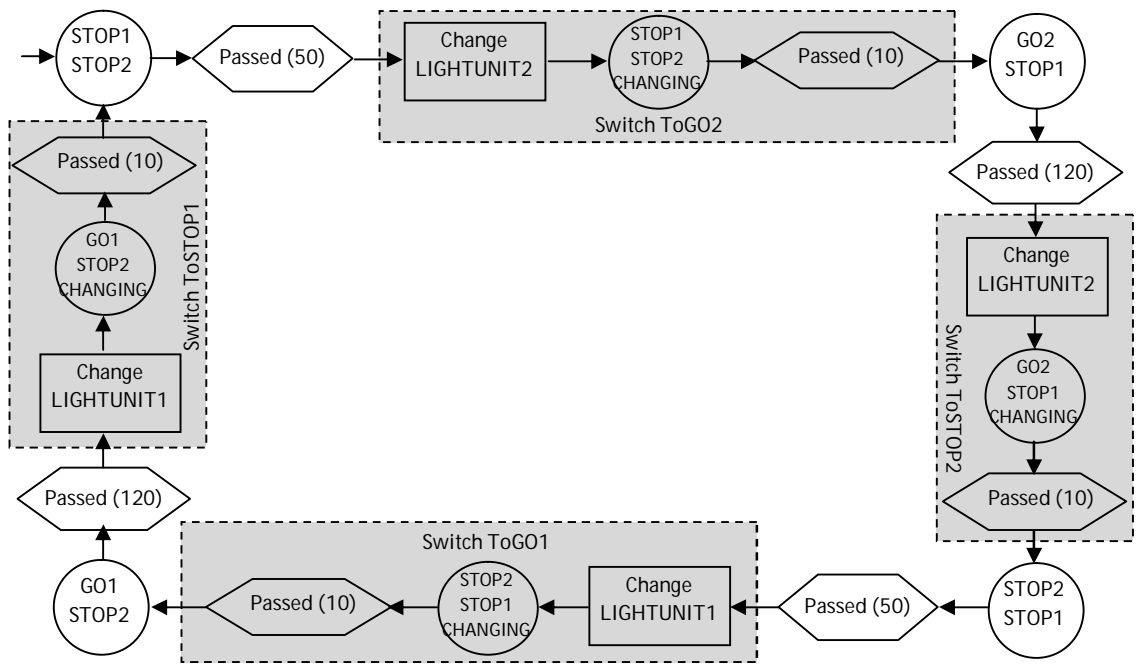

**Fig. 16.** One way traffic light control

We define four boolean controlled functions *GoLight(i)*, *StopLight(i)*, *RPulse(i)*, and *GPulse(i)* for showing the stop light, go light, R pulse, and G pulse of the traffic light units where *i* is LIGHTUNIT1 or LIGHTUNIT2. The GoLight(i) and the StopLight(i) locations are true, if the Go light respectively the Stop light of the traffic light unit *i* is turned on, false otherwise. The RPulse(i) and the GPulse(i) locations are true, if the computer system has
sent to the traffic light unit *i* the pulse to turn on the Stop light respectively the Go light.

To represent the state of the system we define the controlled function phase. Moreover,
The boolean monitored locations *Passed(50)*, *Passed(120)*, and *Passed(10)* indicate when
respectively the Stop time, the Go time, and the change time have passed.

Fig. 17 shows the AsmetaL model.

```
asm oneWayTrafficLightControl

import StandardLibrary
import LTLlibrary

signature:
    enum domain LightUnit = {LIGHTUNIT1 | LIGHTUNIT2}
    enum domain PhaseDomain = { STOP1STOP2 | GO2STOP1 | STOP2STOP1 | GO1STOP2 |
                   STOP1STOP2CHANGING | GO2STOP1CHANGING | STOP2STOP1CHANGING | GO1STOP2CHANGING }
    domain Intervals subsetof Integer
    dynamic controlled phase: PhaseDomain
    dynamic controlled stopLight: LightUnit -> Boolean
    dynamic controlled goLight: LightUnit -> Boolean
    dynamic monitored passed: Intervals -> Boolean
    dynamic controlled rPulse: LightUnit -> Boolean
    dynamic controlled gPulse: LightUnit -> Boolean

definitions:
    domain Intervals = {10, 50, 120}

    rule r_emit($p in Boolean) =
        $p:=true

    rule r_switchLightUnit($l in LightUnit) =
        par
            r_emit[rPulse($l)]
            r_emit[gPulse($l)]
        endpar

    rule r_stop1stop2_to_stop1stop2changing =
        if(phase=STOP1STOP2) then
            if(passed(50)) then
                par
                    r_switchLightUnit[LIGHTUNIT2]
                    phase:=STOP1STOP2CHANGING
                endpar
            endif
        endif

    rule r_stop1stop2changing_to_go2stop1 =
        if(phase=STOP1STOP2CHANGING) then
            if(passed(10)) then
                phase:=GO2STOP1
            endif
        endif

    rule r_go2stop1_to_go2stop1changing =
        if(phase=GO2STOP1) then
            if(passed(120)) then
                par
                    r_switchLightUnit[LIGHTUNIT2]
                    phase:=GO2STOP1CHANGING
                endpar
            endif
        endif

    rule r_go2stop1changing_to_stop2stop1 =
        if(phase=GO2STOP1CHANGING) then
            if(passed(10)) then
                phase:=STOP2STOP1
```

```
            endif
        endif

rule r_stop2stop1_to_stop2stop1changing =
    if(phase=STOP2STOP1) then
        if(passed(50)) then
            par
                r_switchLightUnit[LIGHTUNIT1]
                phase:=STOP2STOP1CHANGING
            endpar
        endif
    endif

rule r_stop2stop1changing_to_go1stop2 =
    if(phase=STOP2STOP1CHANGING) then
        if(passed(10)) then
            phase:=GO1STOP2
        endif
    endif

rule r_go1stop2_to_go1stop2changing =
    if(phase=GO1STOP2) then
        if(passed(120)) then
            par
                r_switchLightUnit[LIGHTUNIT1]
                phase:=GO1STOP2CHANGING
            endpar
        endif
    endif

rule r_go1stop2changing_to_stop1stop2 =
    if(phase=GO1STOP2CHANGING) then
        if(passed(10)) then
            phase:=STOP1STOP2
        endif
    endif

rule r_pulses =
    forall $l in LightUnit with true do
        par
            if(gPulse($l)) then
                par
                    goLight($l) := not(goLight($l))
                    gPulse($l) := false
                endpar
            endif
            if(rPulse($l)) then
                par
                    stopLight($l) := not(stopLight($l))
                    rPulse($l) := false
                endpar
            endif
        endpar

//liveness: correctness of the transitions between states
LTLSPEC NAME ltl_stop1stop2_to_stop1stop2changing:= g((phase=STOP1STOP2 and passed(50))
                                                implies f(phase=STOP1STOP2CHANGING))
LTLSPEC NAME ltl_stop1stop2changing_to_go2stop1:= g((phase=STOP1STOP2CHANGING and passed(10))
                                                implies f(phase=GO2STOP1))
LTLSPEC NAME ltl_go2stop1_to_go2stop1changing:= g((phase=GO2STOP1 and passed(120))
                                                implies f(phase=GO2STOP1CHANGING))
LTLSPEC NAME ltl_go2stop1changing_to_stop2stop1:= g((phase=GO2STOP1CHANGING and passed(10))
                                                implies f(phase=STOP2STOP1))
LTLSPEC NAME ltl_stop2stop1_to_stop2stop1changing:= g((phase=STOP2STOP1 and passed(50))
                                                implies f(phase=STOP2STOP1CHANGING))
LTLSPEC NAME ltl_stop2stop1changing_to_go1stop2:= g((phase=STOP2STOP1CHANGING and passed(10))
                                                implies f(phase=GO1STOP2))
LTLSPEC NAME ltl_go1stop2_to_go1stop2changing:= g((phase=GO1STOP2 and passed(120))
                                                implies f(phase=GO1STOP2CHANGING))
LTLSPEC NAME ltl_go1stop2changing_to_stop1stop2:= g((phase=GO1STOP2CHANGING and passed(10))
                                                implies f(phase=STOP1STOP2))
```

```
//properties on the relationship between the phase and traffic lights
LTLSPEC NAME ltl_phase:= g(((phase=STOP1STOP2 or phase=STOP2STOP1) and
                    stopLight(LIGHTUNIT1) and not(goLight(LIGHTUNIT1)) and
                    stopLight(LIGHTUNIT2) and not(goLight(LIGHTUNIT2)))
                    or
                   (phase=GO2STOP1 and not(stopLight(LIGHTUNIT2)) and goLight(LIGHTUNIT2)
                    and stopLight(LIGHTUNIT1) and not(goLight(LIGHTUNIT1)))
                    or
                   (phase=GO1STOP2 and not(stopLight(LIGHTUNIT1)) and goLight(LIGHTUNIT1)
                    and stopLight(LIGHTUNIT2) and not(goLight(LIGHTUNIT2)))
                    or
                   (phase=STOP1STOP2CHANGING or phase=GO2STOP1CHANGING or
                    phase=STOP2STOP1CHANGING or phase=GO1STOP2CHANGING))

//the Go light of both traffic light units cannot on at the same time
LTLSPEC NAME ltl_goLights:= g(not(goLight(LIGHTUNIT1) and goLight(LIGHTUNIT2)))

//this property is equivalent to the previous one
LTLSPEC NAME ltl_Lights:= g((goLight(LIGHTUNIT2)  and  stopLight(LIGHTUNIT1)) xor
                    (goLight(LIGHTUNIT1)  and  stopLight(LIGHTUNIT2)) xor
                    (stopLight(LIGHTUNIT2)  and  stopLight(LIGHTUNIT1)))

//either the stop light or the go light is turned on (not both)
LTLSPEC NAME ltl_Light1:=  g(goLight(LIGHTUNIT1) xor stopLight(LIGHTUNIT1))
LTLSPEC NAME ltl_Light2:=  g(goLight(LIGHTUNIT2) xor stopLight(LIGHTUNIT2))

main rule r_Main =
    par
        r_stop1stop2_to_stop1stop2changing[]
        r_stop1stop2changing_to_go2stop1[]
        r_go2stop1_to_go2stop1changing[]
        r_go2stop1changing_to_stop2stop1[]
        r_stop2stop1_to_stop2stop1changing[]
        r_stop2stop1changing_to_go1stop2[]
        r_go1stop2_to_go1stop2changing[]
        r_go1stop2changing_to_stop1stop2[]
        r_pulses[]
    endpar

default init s0:
    function stopLight($l in LightUnit) = true
    function goLight($l in LightUnit) = false
    function phase = STOP1STOP2
    function rPulse($l in LightUnit) = false
    function gPulse($l in LightUnit) = false
```

**Fig. 17.** One way traffic light control: AsmetaL model

Let's see the main rule. It consists of nine rules which model the behaviour of the system.

- *r_stop1stop2_to_ stop1stop2changing*,
- *r_ stop1stop2changing _to_ go2stop1*,
- *r_ go2stop1_to_ go2stop1changing*,
- *r_ go2stop1changing _to_ stop2stop1*,
- *r_ stop2stop1_to_ stop2stop1changing*,
- *r_ stop2stop1changing _to_ go1stop2*,
- *r_ go1stop2_to_ go1stop2changing*,
- *r_ go1stop2changing _to_ stop1stop2*,
- *r_pulses*

*r_stop1stop2_to_ stop1stop2changing* rule says, if the state of the system is STOP1STOP2 (*Phase== STOP1STOP2*) and the stop period is passed (*Passed(50)*), then the computer system changes the LIGHTUNIT2 with the macro call rule *r_switchLightUnit[LIGHTUNIT2]* and the system enters phase STOP1STOP2CHANGING (*Phase:= STOP1STOP2CHANGING*).

The rule *r_switchLightUnit($i in LightUnit)* sends the R and G pulses to the traffic light unit *$i*, with two macro call rules *r_emit[RPulse($i)]* and *r_emit[GPulse($i)]*, where LightUnit is an enum domain consists of two traffic light units, LIGHTUNIT1 and LIGHTUNIT2.

To handle the received pulses, we define the rule *r_pulses*; when it detects R or G pulse for the traffic light *i*, it complements the Stop light respectively the Go light of *i*.

We also add some properties we want to check.

To check the correctness of the transitions between states we declare the liveness properties:

*LTLSPEC NAME ltl_stop1stop2_to_stop1stop2changing:= g((phase=STOP1STOP2 and passed(50))*
*implies f(phase=STOP1STOP2CHANGING))*
*LTLSPEC NAME ltl_stop1stop2changing_to_go2stop1:= g((phase=STOP1STOP2CHANGING and passed(10))*
*implies f(phase=GO2STOP1))*
*LTLSPEC NAME ltl_go2stop1_to_go2stop1changing:= g((phase=GO2STOP1 and passed(120))*
*implies f(phase=GO2STOP1CHANGING))*
*LTLSPEC NAME ltl_go2stop1changing_to_stop2stop1:= g((phase=GO2STOP1CHANGING and passed(10))*
*implies f(phase=STOP2STOP1))*
*LTLSPEC NAME ltl_stop2stop1_to_stop2stop1changing:= g((phase=STOP2STOP1 and passed(50))*
*implies f(phase=STOP2STOP1CHANGING))*
*LTLSPEC NAME ltl_stop2stop1changing_to_go1stop2:= g((phase=STOP2STOP1CHANGING and passed(10))*
*implies f(phase=GO1STOP2))*
*LTLSPEC NAME ltl_go1stop2_to_go1stop2changing:= g((phase=GO1STOP2 and passed(120))*
*implies f(phase=GO1STOP2CHANGING))*
*LTLSPEC NAME ltl_go1stop2changing_to_stop1stop2:= g((phase=GO1STOP2CHANGING and passed(10))*
*implies f(phase=STOP1STOP2))*

Safety property

*LTLSPEC NAME ltl_phase:= g(*
*((phase=STOP1STOP2 or phase=STOP2STOP1) and*
*stopLight(LIGHTUNIT1) and not(goLight(LIGHTUNIT1)) and*
*stopLight(LIGHTUNIT2) and not(goLight(LIGHTUNIT2)))*
*or*
*(phase=GO2STOP1 and not(stopLight(LIGHTUNIT2)) and goLight(LIGHTUNIT2)*
*and stopLight(LIGHTUNIT1) and not(goLight(LIGHTUNIT1)))*
*or*
*(phase=GO1STOP2 and not(stopLight(LIGHTUNIT1)) and goLight(LIGHTUNIT1)*
*and stopLight(LIGHTUNIT2) and not(goLight(LIGHTUNIT2)))*
*or*
*(phase=STOP1STOP2CHANGING or phase=GO2STOP1CHANGING or*
*phase=STOP2STOP1CHANGING or phase=GO1STOP2CHANGING))*

checks that, in each state, the relation between phase and traffic lights is correct. For example, if the phase is STOP1STOP2 or STOP2STOP1, both traffic light units must show stop light.

Safety property

*LTLSPEC NAME ltl_goLights:= g(not(goLight(LIGHTUNIT1) and goLight(LIGHTUNIT2)))*

says that, the Go light of both traffic light units cannot on at the same time.

Safety property

*LTLSPEC NAME ltl_Lights:= g((goLight(LIGHTUNIT2) and stopLight(LIGHTUNIT1)) xor*

*(goLight(LIGHTUNIT1) and stopLight(LIGHTUNIT2)) xor*
*(stopLight(LIGHTUNIT2) and stopLight(LIGHTUNIT1)))*

is equivalent to the previous one. It says that, if a traffic light unit shows Go, another must show Stop or both units must show Stop.

Safety properties

*LTLSPEC NAME ltl_Light1:= g(goLight(LIGHTUNIT1) xor stopLight(LIGHTUNIT1))*
*LTLSPEC NAME ltl_Light2:= g(goLight(LIGHTUNIT2) xor stopLight(LIGHTUNIT2))*

check that, in each state, in each traffic light, either the Stop light or the Go light is turned on, not both.

Fig. 18 is the equivalent BIR code of the AsmetaL model represented in Fig. 17. We simplify the BIR model by the merging of all locations which have the same BIR guarded commands. We also remove the unused locations.

```
system oneWayTrafficLightControl {

    fun ltl_stop1stop2_to_stop1stop2changing() returns boolean =
        LTL.temporalProperty(
            Property.createObservableDictionary(
                Property.createObservableKey("P1", phase==PhaseDomain.STOP1STOP2),
                Property.createObservableKey("P2", passed_50),
                Property.createObservableKey("P3", phase==PhaseDomain.STOP1STOP2CHANGING)),
            LTL.always(
                LTL.implication(
                    LTL.conjunction(LTL.prop("P1"),LTL.prop("P2")),
                    LTL.eventually(LTL.prop("P3"))
                )
            )
        );

    .
    . //other seven liveness properties are similar to the above function.
    .

    fun ltl_phase() returns boolean =
        LTL.temporalProperty(
            Property.createObservableDictionary(
                Property.createObservableKey("P1", phase==PhaseDomain.STOP1STOP2),
                Property.createObservableKey("P2", phase==PhaseDomain.STOP2STOP1),
                Property.createObservableKey("P3", stopLight_LIGHTUNIT1),
                Property.createObservableKey("P4", goLight_LIGHTUNIT1),
                Property.createObservableKey("P5", stopLight_LIGHTUNIT2),
                Property.createObservableKey("P6", goLight_LIGHTUNIT2),
                Property.createObservableKey("P7", phase==PhaseDomain.GO2STOP1),
                Property.createObservableKey("P8", phase==PhaseDomain.GO1STOP2),
                Property.createObservableKey("P9", phase==PhaseDomain.STOP1STOP2CHANGING),
                Property.createObservableKey("P10", phase==PhaseDomain.GO2STOP1CHANGING),
                Property.createObservableKey("P11", phase==PhaseDomain.STOP2STOP1CHANGING),
                Property.createObservableKey("P12", phase==PhaseDomain.GO1STOP2CHANGING)),
            LTL.always(
                LTL.disjunction(
                    LTL.disjunction(
                        LTL.disjunction(
                            LTL.conjunction(
                                LTL.conjunction(
                                    LTL.conjunction(
                                        LTL.disjunction(LTL.prop("P1"),LTL.prop("P2")),LTL.prop("P3")
                                    ),LTL.negation(LTL.prop("P4"))
                                ),LTL.prop("P5")
                            ),LTL.negation(LTL.prop("P6"))
                        ),LTL.conjunction(
                            LTL.conjunction(
                                LTL.conjunction(
```

```
                                LTL.conjunction(
                                    LTL.prop("P7"),LTL.negation(LTL.prop("P5"))
                                ),LTL.prop("P6")
                            ),LTL.prop("P3")
                        ),LTL.negation(LTL.prop("P4")))
                ),LTL.conjunction(
                        LTL.conjunction(
                            LTL.conjunction(
                                LTL.conjunction(
                                    LTL.prop("P8"),LTL.negation(LTL.prop("P3"))
                                ),LTL.prop("P4")
                            ),LTL.prop("P5")
                        ),LTL.negation(LTL.prop("P6")))
                ),LTL.disjunction(
                        LTL.disjunction(
                            LTL.disjunction(LTL.prop("P9"),LTL.prop("P10")),LTL.prop("P11")
                        ),LTL.prop("P12"))))
    );

fun ltl_goLights() returns boolean =
    LTL.temporalProperty(
        Property.createObservableDictionary(
            Property.createObservableKey("P1", goLight_LIGHTUNIT1),
            Property.createObservableKey("P2", goLight_LIGHTUNIT2)),
        LTL.always(LTL.negation(LTL.conjunction(LTL.prop("P1"),LTL.prop("P2"))))
    );
.
.
.

enum LightUnit {LIGHTUNIT1 , LIGHTUNIT2}
enum PhaseDomain { STOP1STOP2, GO2STOP1, STOP2STOP1, GO1STOP2, STOP1STOP2CHANGING,
                   GO2STOP1CHANGING, STOP2STOP1CHANGING, GO1STOP2CHANGING }
typealias Intervals int;
PhaseDomain phase;//controlled
boolean stopLight_LIGHTUNIT1;//controlled
boolean stopLight_LIGHTUNIT2;//controlled
boolean goLight_LIGHTUNIT1;//controlled
boolean goLight_LIGHTUNIT2;//controlled
boolean passed_10;//monitored
boolean passed_50;//monitored
boolean passed_120;//monitored
boolean rPulse_LIGHTUNIT1;//controlled
boolean rPulse_LIGHTUNIT2;//controlled
boolean gPulse_LIGHTUNIT1;//controlled
boolean gPulse_LIGHTUNIT2;//controlled

thread passed_10_monitored(){
    loc loc0:
        do {passed_10:=true;} goto loc0;
        do {passed_10:=false;} goto loc0;
}
thread passed_50_monitored(){
    loc loc0:
        do {passed_50:=true;} goto loc0;
        do {passed_50:=false;} goto loc0;
}
thread passed_120_monitored(){
    loc loc0:
        do {passed_120:=true;} goto loc0;
        do {passed_120:=false;} goto loc0;
}

active thread MAIN(){
    loc loc0://initialization
        do{
            stopLight_LIGHTUNIT1:=true;
            stopLight_LIGHTUNIT2:=true;
            goLight_LIGHTUNIT1:=false;
            goLight_LIGHTUNIT2:=false;
            phase:=PhaseDomain.STOP1STOP2;
```

```
        rPulse_LIGHTUNIT1:=false;
        rPulse_LIGHTUNIT2:=false;
        gPulse_LIGHTUNIT1:=false;
        gPulse_LIGHTUNIT2:=false;

        //monitored functions
        start passed_10_monitored();
        start passed_50_monitored();
        start passed_120_monitored();
    }goto loc1;

//rules
//stop1stop2 -> stop1stop2changing
loc loc1:
    when (phase==PhaseDomain.STOP1STOP2 && passed_50) do invisible {
        rPulse_LIGHTUNIT2:=true;
        gPulse_LIGHTUNIT2:=true;
        phase:=PhaseDomain.STOP1STOP2CHANGING;
    }goto loc2_C1;
    when !(phase==PhaseDomain.STOP1STOP2 && passed_50) do invisible {
    } goto loc2_N1;

//stop1stop2changing -> go2stop1
loc loc2_C1:
    do invisible{}goto loc3;
loc loc2_N1:
    when (phase==PhaseDomain.STOP1STOP2CHANGING && passed_10) do invisible {
        phase:=PhaseDomain.GO2STOP1;
    }goto loc2_C1;
    when !(phase==PhaseDomain.STOP1STOP2CHANGING && passed_10) do invisible {
    } goto loc2_N1_N2;

//go2stop1 -> go2stop1changing
loc loc2_N1_N2:
    when (phase==PhaseDomain.GO2STOP1 && passed_120) do invisible {
        rPulse_LIGHTUNIT2:=true;
        gPulse_LIGHTUNIT2:=true;
        phase:=PhaseDomain.GO2STOP1CHANGING;
    }goto loc2_C1;
    when !(phase==PhaseDomain.GO2STOP1 && passed_120) do invisible {
    } goto loc2_N1_N2_N3;

//go2stop1changing -> stop2stop1
loc loc2_N1_N2_N3:
    when (phase==PhaseDomain.GO2STOP1CHANGING && passed_10) do invisible {
        phase:=PhaseDomain.STOP2STOP1;
    }goto loc2_C1;
    when !(phase==PhaseDomain.GO2STOP1CHANGING && passed_10) do invisible {
    } goto loc2_N1_N2_N3_N4;

//stop2stop1 -> stop2stop1changing
loc loc2_N1_N2_N3_N4:
    when (phase==PhaseDomain.STOP2STOP1 && passed_50) do invisible {
        rPulse_LIGHTUNIT1:=true;
        gPulse_LIGHTUNIT1:=true;
        phase:=PhaseDomain.STOP2STOP1CHANGING;
    }goto loc2_C1;
    when !(phase==PhaseDomain.STOP2STOP1 && passed_50) do invisible {
    } goto loc2_N1_N2_N3_N4_N5;

//stop2stop1changing -> go1stop2
loc loc2_N1_N2_N3_N4_N5:
    when (phase==PhaseDomain.STOP2STOP1CHANGING && passed_10) do invisible {
        phase:=PhaseDomain.GO1STOP2;
    }goto loc2_C1;
    when !(phase==PhaseDomain.STOP2STOP1CHANGING && passed_10) do invisible {
    } goto loc2_N1_N2_N3_N4_N5_N6;

//go1stop2 -> go1stop2changing
loc loc2_N1_N2_N3_N4_N5_N6:
    when (phase==PhaseDomain.GO1STOP2 && passed_120) do invisible {
```

```
                rPulse_LIGHTUNIT1:=true;
                gPulse_LIGHTUNIT1:=true;
                phase:=PhaseDomain.GO1STOP2CHANGING;
            }goto loc2_C1;
            when !(phase==PhaseDomain.GO1STOP2 && passed_120) do invisible {
            } goto loc2_N1_N2_N3_N4_N5_N6_N7;

    //go1stop2changing -> stop1stop2
    loc loc2_N1_N2_N3_N4_N5_N6_N7:
            when (phase==PhaseDomain.GO1STOP2CHANGING && passed_10) do invisible {
                phase:=PhaseDomain.STOP1STOP2;
            }goto loc2_C1;
            when !(phase==PhaseDomain.GO1STOP2CHANGING && passed_10) do invisible {
            } goto loc3;

    loc loc3:
            when (gPulse_LIGHTUNIT1) do invisible {
                goLight_LIGHTUNIT1:=!(goLight_LIGHTUNIT1);
                gPulse_LIGHTUNIT1:=false;
            }goto loc3_C1;
            when !(gPulse_LIGHTUNIT1) do invisible {} goto loc3_C1;

    loc loc3_C1:
            when (rPulse_LIGHTUNIT1) do invisible {
                stopLight_LIGHTUNIT1:=!(stopLight_LIGHTUNIT1);
                rPulse_LIGHTUNIT1:=false;
            }goto loc3_C1_C2;
            when !(rPulse_LIGHTUNIT1) do invisible {} goto loc3_C1_C2;

    loc loc3_C1_C2:
            when (gPulse_LIGHTUNIT2) do invisible {
                goLight_LIGHTUNIT2:=!(goLight_LIGHTUNIT2);
                gPulse_LIGHTUNIT2:=false;
            }goto loc3_C1_C2_C3;
            when !(gPulse_LIGHTUNIT2) do invisible {} goto loc3_C1_C2_C3;

    loc loc3_C1_C2_C3:
            when (rPulse_LIGHTUNIT2) do invisible {
                stopLight_LIGHTUNIT2:=!(stopLight_LIGHTUNIT2);
                rPulse_LIGHTUNIT2:=false;
            }goto endloc;
            when !(rPulse_LIGHTUNIT2) do invisible {} goto endloc;

    loc endloc:
            do {/*Bogor creates a state*/} goto loc1;
    }
}
```

**Fig. 18.** One way traffic light control: BIR model

By now, we execute the BIR code in order to check the deadlock and correctness of the desired properties.

user@linux-cdhm:~> asm2bogor.sh oneWayTrafficLightControl.asm
Bogor v.1.2 (build <version>)
(c) Copyright by Kansas State University

Web: http://bogor.projects.cis.ksu.edu

Transitions: 1137, States: 65, Matched States: 385, Max Depth: 210, Errors found: 0, Used Memory: 0MB
** oneWayTrafficLightControl is not in DEADLOCK
**LTLSPEC NAME ltl_stop1stop2_to_stop1stop2changing:= g((phase=STOP1STOP2 and passed(50)) implies f(phase=STOP1STOP2CHANGING)) is true
**LTLSPEC NAME ltl_stop1stop2changing_to_go2stop1:= g((phase=STOP1STOP2CHANGING and passed(10)) implies f(phase=GO2STOP1)) is true
**LTLSPEC NAME ltl_go2stop1_to_go2stop1changing:= g((phase=GO2STOP1 and passed(120)) implies f(phase=GO2STOP1CHANGING)) is true

**LTLSPEC NAME ltl_go2stop1changing_to_stop2stop1:= g((phase=GO2STOP1CHANGING and passed(10)) implies f(phase=STOP2STOP1)) is true

**LTLSPEC NAME ltl_stop2stop1_to_stop2stop1changing:= g((phase=STOP2STOP1 and passed(50)) implies f(phase=STOP2STOP1CHANGING)) is true

**LTLSPEC NAME ltl_stop2stop1changing_to_go1stop2:= g((phase=STOP2STOP1CHANGING and passed(10)) implies f(phase=GO1STOP2)) is true

**LTLSPEC NAME ltl_go1stop2_to_go1stop2changing:= g((phase=GO1STOP2 and passed(120)) implies f(phase=GO1STOP2CHANGING)) is true

**LTLSPEC NAME ltl_go1stop2changing_to_stop1stop2:= g((phase=GO1STOP2CHANGING and passed(10)) implies f(phase=STOP1STOP2)) is true

**LTLSPEC NAME ltl_phase:= g(((phase=STOP1STOP2 or phase=STOP2STOP1) and stopLight(LIGHTUNIT1) and not(goLight(LIGHTUNIT1)) and stopLight(LIGHTUNIT2) and not(goLight(LIGHTUNIT2))) or (phase=GO2STOP1 and not(stopLight(LIGHTUNIT2)) and goLight(LIGHTUNIT2) and stopLight(LIGHTUNIT1) and not(goLight(LIGHTUNIT1))) or (phase=GO1STOP2 and not(stopLight(LIGHTUNIT1)) and goLight(LIGHTUNIT1) and stopLight(LIGHTUNIT2) and not(goLight(LIGHTUNIT2))) or (phase=STOP1STOP2CHANGING or phase=GO2STOP1CHANGING or phase=STOP2STOP1CHANGING or phase=GO1STOP2CHANGING)) is true

**LTLSPEC NAME ltl_goLights:= g(not(goLight(LIGHTUNIT1) and goLight(LIGHTUNIT2))) is true

**LTLSPEC NAME ltl_Lights:= g((goLight(LIGHTUNIT1)) xor (goLight(LIGHTUNIT1) and stopLight(LIGHTUNIT2)) xor (stopLight(LIGHTUNIT2) and stopLight(LIGHTUNIT1))) is true

**LTLSPEC NAME ltl_Light1:= g(goLight(LIGHTUNIT1) xor stopLight(LIGHTUNIT1)) is true

**LTLSPEC NAME ltl_Light2:= g(goLight(LIGHTUNIT2) xor stopLight(LIGHTUNIT2)) is true

Done!

### 7.3.    Dining philosophers

Dining philosopher problem is an example for a distributed system of communicating processes. So, we model it by means of a multi-agent ASM that each agent describes a philosopher: *n* philosophers sit around a circular table. Also, there only exist *n* fork. Each fork is placed between each pair of philosophers.

The AsmetaL code of the model is presented in Fig 19.

```
asm diningPhilosophers

import StandardLibrary
import LTLlibrary

signature:
    domain Philosophers subsetof Agent
    abstract domain Fork
    monitored hungry: Philosophers -> Boolean
    controlled eating: Philosophers -> Boolean
    static right_fork: Philosophers -> Fork
    static left_fork: Philosophers -> Fork
    controlled owner: Fork -> Philosophers
    static phil_1: Philosophers
    static phil_2: Philosophers
    static phil_3: Philosophers
    static phil_4: Philosophers
    static phil_5: Philosophers
    static fork_1: Fork
    static fork_2: Fork
    static fork_3: Fork
    static fork_4: Fork
    static fork_5: Fork

definitions:
    function right_fork($a in Philosophers) =
        switch($a)
            case phil_1: fork_2
            case phil_2: fork_3
            case phil_3: fork_4
            case phil_4: fork_5
```

```
            otherwise    fork_1
        endswitch

function left_fork($a in Philosophers) =
    switch($a)
        case phil_1: fork_1
        case phil_2: fork_2
        case phil_3: fork_3
        case phil_4: fork_4
        otherwise    fork_5
    endswitch

macro rule r_Eat =
    if (hungry(self)) then
        if( isUndef(owner(left_fork(self))) and isUndef(owner(right_fork(self))) ) then
            par
                owner(left_fork(self)) := self
                owner(right_fork(self)) := self
                eating(self) := true
            endpar
        endif
    endif

macro rule r_Think =
    if ( not hungry(self)) then
        if eating(self) and (owner(left_fork(self))=self)and(owner(right_fork(self))=self) then
            par
                owner(left_fork(self)) := undef
                owner(right_fork(self)) := undef
                eating(self) := false
            endpar
        endif
    endif

macro rule r_Philo =
    par
        r_Eat[]
        r_Think[]
    endpar

LTLSPEC NAME ltl_HungryToEatingPhil1:= g(
            (hungry(phil_1) and isUndef(owner(fork_1)) and isUndef(owner(fork_2)))
            implies (eating(phil_1) iff (owner(fork_1=phil_1 and owner(fork_2)=phil_1))
                )
LTLSPEC NAME ltl_HungryToEatingPhil2:= g(
            (hungry(phil_2) and isUndef(owner(fork_2)) and isUndef(owner(fork_3)))
            implies (eating(phil_2) iff (owner(fork_1=phil_2 and owner(fork_2)=phil_2))
                )
LTLSPEC NAME ltl_HungryToEatingPhil3:= g(
            (hungry(phil_3) and isUndef(owner(fork_3)) and isUndef(owner(fork_4)))
            implies (eating(phil_3) iff (owner(fork_1=phil_3 and owner(fork_2)=phil_3))
                )
LTLSPEC NAME ltl_HungryToEatingPhil4:= g(
            (hungry(phil_4) and isUndef(owner(fork_4)) and isUndef(owner(fork_5)))
            implies (eating(phil_4) iff (owner(fork_1=phil_4 and owner(fork_2)=phil_4))
                )
LTLSPEC NAME ltl_HungryToEatingPhil5:= g(
            (hungry(phil_5) and isUndef(owner(fork_5)) and isUndef(owner(fork_1)))
            implies (eating(phil_5) iff (owner(fork_1=phil_5 and owner(fork_2)=phil_5))
                )

LTLSPEC NAME ltl_chkFork1:= g(not(owner(fork_1)=phil_1 and owner(fork_1)=phil_5))
LTLSPEC NAME ltl_chkFork2:= g(not(owner(fork_2)=phil_1 and owner(fork_2)=phil_2))
LTLSPEC NAME ltl_chkFork3:= g(not(owner(fork_3)=phil_2 and owner(fork_3)=phil_3))
LTLSPEC NAME ltl_chkFork4:= g(not(owner(fork_4)=phil_3 and owner(fork_4)=phil_4))
LTLSPEC NAME ltl_chkFork5:= g(not(owner(fork_5)=phil_4 and owner(fork_5)=phil_5))

LTLSPEC NAME ltl_eatingOfNeighbours:= g(not(eating(phil_1) and eating(phil_2)) or
                              not(eating(phil_2) and eating(phil_3)) or
                              not(eating(phil_3) and eating(phil_4)) or
                              not(eating(phil_4) and eating(phil_5)) or
```

```
                    not(eating(phil_5) and eating(phil_1)))

    main rule r_choose_philo = choose $p in Philosophers with true do program($p)

default init s0:
    function eating ($p in Philosophers)= false
    function owner ($f in Fork) = undef
    agent Philosophers: r_Philo[]
```

**Fig. 18.** Dining philosophers problem: AsmetaL model

We define two domains, Philosophers and Forks, and two static functions, left_fork and right_fork for modeling the problem. Theses static functions express a fixed relation between the left and the right fork of a philosopher. Since the value of think, hungry, eating, and owner is changed by philosophers during a run, we introduce them as controlled terms. For example, a philosopher could get hungry, so he/she will change the value of his/her think attribute and his/her hungry attribute.

Each philosopher is in a think, hungry, or eating cycle. At the beginning he/she thinks. Then he/she gets hungry. If the philosopher can pick up both his/her right fork and left fork, he/she eats for some times.

Finally, whenever he/she is not hungry but still eating then he/she stop eating and put the forks down and starts to think. Since, the behaviour of each philosopher is the same, we model his/her behaviour by means of parameterized rules. The parameter of these rules is *self* which ranges over all philosophers.

We also add some properties we want to check.

The five safety properties

*LTLSPEC NAME ltl_HungryToEatingPhil1:= g((hungry(phil_1) and isUndef(owner(fork_1)) and*
  *isUndef(owner(fork_2))) implies (eating(phil_1) iff (owner(fork_1)=phil_1 and owner(fork_2)=phil_1)))*
*LTLSPEC NAME ltl_HungryToEatingPhil2:= g((hungry(phil_2) and isUndef(owner(fork_2)) and*
  *isUndef(owner(fork_3))) implies (eating(phil_2) iff (owner(fork_1)=phil_2 and owner(fork_2)=phil_2)))*
*LTLSPEC NAME ltl_HungryToEatingPhil3:= g((hungry(phil_3) and isUndef(owner(fork_3)) and*
  *isUndef(owner(fork_4))) implies (eating(phil_3) iff (owner(fork_1)=phil_3 and owner(fork_2)=phil_3)))*
*LTLSPEC NAME ltl_HungryToEatingPhil4:= g((hungry(phil_4) and isUndef(owner(fork_4)) and*
  *isUndef(owner(fork_5))) implies (eating(phil_4) iff (owner(fork_1)=phil_4 and owner(fork_2)=phil_4)))*
*LTLSPEC NAME ltl_HungryToEatingPhil5:= g((hungry(phil_5) and isUndef(owner(fork_5)) and*
  *isUndef(owner(fork_1))) implies (eating(phil_5) iff (owner(fork_1)=phil_5 and owner(fork_2)=phil_5)))*

are true, if in every possible execution there is a state in a path in which *phil_i* is hungry and his/her forks are free, then *phil_i* can eat dinner if and only if he/she pick up the forks.

The safety properties

*LTLSPEC NAME ltl_chkFork1:= g(not(owner(fork_1)=phil_1 and owner(fork_1)=phil_5))*
*LTLSPEC NAME ltl_chkFork2:= g(not(owner(fork_2)=phil_1 and owner(fork_2)=phil_2))*
*LTLSPEC NAME ltl_chkFork3:= g(not(owner(fork_3)=phil_2 and owner(fork_3)=phil_3))*
*LTLSPEC NAME ltl_chkFork4:= g(not(owner(fork_4)=phil_3 and owner(fork_4)=phil_4))*
*LTLSPEC NAME ltl_chkFork5:= g(not(owner(fork_5)=phil_4 and owner(fork_5)=phil_5))*

check that, the two neighbour philosopher cannot pick up the fork which is placed between them. An equivalent safety property to the properties is

*LTLSPEC NAME ltl_eatingOfNeighbours:= g(not(eating(phil_1) and eating(phil_2)) or*
  *not(eating(phil_2) and eating(phil_3)) or*
  *not(eating(phil_3) and eating(phil_4)) or*
  *not(eating(phil_4) and eating(phil_5)) or*
  *not(eating(phil_5) and eating(phil_1)))*

that expresses that, in each state of the ASM model, there should be no state in which the neighbour philosophers are eating together.

Fig. 19 is the equivalent BIR code of the AsmetaL model represented in Fig. 18.

```
system diningPhilosophers {

    fun ltl_HungryToEatingPhil1() returns boolean =
        LTL.temporalProperty(
            Property.createObservableDictionary(
                Property.createObservableKey("P1", phil_1.hungry),
                Property.createObservableKey("P2", fork_1.owner==null),
                Property.createObservableKey("P3", fork_2.owner==null),
                Property.createObservableKey("P4", phil_1.eating),
                Property.createObservableKey("P5", fork_1.owner==phil_1),
                Property.createObservableKey("P6", fork_2.owner==phil_1)),
            LTL.always(LTL.implication(
                LTL.conjunction(LTL.conjunction(LTL.prop("P1"),LTL.prop("P2")),LTL.prop("P3")),
                LTL.equivalence(LTL.prop("P4"),LTL.conjunction(LTL.prop("P5"),LTL.prop("P6")))
            ))
        );
    .
    .
    .
    fun ltl_chkFork1() returns boolean =
        LTL.temporalProperty(
            Property.createObservableDictionary(
                Property.createObservableKey("P1", fork_1.owner==phil_1),
                Property.createObservableKey("P2", fork_1.owner==phil_5)),
            LTL.always(LTL.negation(LTL.conjunction(LTL.prop("P1"),LTL.prop("P2"))))
        );
    .
    .
    .
    fun ltl_eatingOfNeighbours() returns boolean =
        LTL.temporalProperty(
            Property.createObservableDictionary(
                Property.createObservableKey("P1", phil_1.eating),
                Property.createObservableKey("P2", phil_2.eating),
                Property.createObservableKey("P3", phil_3.eating),
                Property.createObservableKey("P4", phil_4.eating),
                Property.createObservableKey("P5", phil_5.eating)),
            LTL.always(
                LTL.disjunction(LTL.disjunction(LTL.disjunction(LTL.disjunction(
                    LTL.negation(LTL.conjunction(LTL.prop("P1"),LTL.prop("P2"))),
                    LTL.negation(LTL.conjunction(LTL.prop("P2"),LTL.prop("P3")))),
                    LTL.negation(LTL.conjunction(LTL.prop("P3"),LTL.prop("P4")))),
                    LTL.negation(LTL.conjunction(LTL.prop("P4"),LTL.prop("P5")))),
                    LTL.negation(LTL.conjunction(LTL.prop("P5"),LTL.prop("P2"))))
            )
        );

    record Philosophers {
        boolean hungry;//monitored
        boolean eating;//controlled
    }
    record Fork {
        Philosophers owner;//controlled
    }
    Philosophers phil_1;//static
    Philosophers phil_2;//static
    Philosophers phil_3;//static
    Philosophers phil_4;//static
    Philosophers phil_5;//static
    Fork fork_1;//static
    Fork fork_2;//static
    Fork fork_3;//static
    Fork fork_4;//static
    Fork fork_5;//static
```

```
fun right_fork(Philosophers a) returns Fork =
    ((a==phil_1?fork_2:(a==phil_2?fork_3:(a==phil_3?fork_4:(a==phil_4?fork_5:fork_1)))));

fun left_fork(Philosophers a) returns Fork =
    ((a==phil_1?fork_1:(a==phil_2?fork_2:(a==phil_3?fork_3:(a==phil_4?fork_4:fork_5)))));

thread phil_1_hungry_monitored() {
    loc loc0 :
        do {phil_1.hungry:=true;} goto loc0;
        do {phil_1.hungry:=false;} goto loc0;
}
thread phil_2_hungry_monitored() {
    loc loc0:
        do {phil_2.hungry:=true;} goto loc0;
        do {phil_2.hungry:=false;} goto loc0;
}
thread phil_3_hungry_monitored() {
    loc loc0:
        do {phil_3.hungry:=true;} goto loc0;
        do {phil_3.hungry:=false;} goto loc0;
}
thread phil_4_hungry_monitored() {
    loc loc0:
        do {phil_4.hungry:=true;} goto loc0;
        do {phil_4.hungry:=false;} goto loc0;
}
thread phil_5_hungry_monitored() {
    loc loc0:
        do {phil_5.hungry:=true;} goto loc0;
        do {phil_5.hungry:=false;} goto loc0;
}

active thread MAIN() {
    loc loc0://initialization
        do {
            phil_1:= new Philosophers;
            phil_2:= new Philosophers;
            phil_3:= new Philosophers;
            phil_4:= new Philosophers;
            phil_5:= new Philosophers;
            fork_1:= new Fork;
            fork_2:= new Fork;
            fork_3:= new Fork;
            fork_4:= new Fork;
            fork_5:= new Fork;
            phil_1.eating:=false;
            phil_2.eating:=false;
            phil_3.eating:=false;
            phil_4.eating:=false;
            phil_5.eating:=false;
            fork_1.owner:=null;
            fork_2.owner:=null;
            fork_3.owner:=null;
            fork_4.owner:=null;
            fork_5.owner:=null;

            //monitored functions
            start phil_1_hungry_monitored();
            start phil_2_hungry_monitored();
            start phil_3_hungry_monitored();
            start phil_4_hungry_monitored();
            start phil_5_hungry_monitored();
        }goto loc1;

    //rules
    loc loc1:
        //philosopher 1
        when (left_fork(phil_1).owner==null && right_fork(phil_1).owner==null && phil_1.hungry)
        do invisible{
            left_fork(phil_1).owner:=phil_1;
            right_fork(phil_1).owner:=phil_1;
```

```
      phil_1.eating:=true;
   }goto loc2_phil_1_C1;
   when !(left_fork(phil_1).owner==null && right_fork(phil_1).owner==null && phil_1.hungry)
   do invisible {} goto loc2_phil_1_N1;

   //philosopher 2
   when (left_fork(phil_2).owner==null && right_fork(phil_2).owner==null && phil_2.hungry)
   do invisible{
      left_fork(phil_2).owner:=phil_2;
      right_fork(phil_2).owner:=phil_2;
      phil_2.eating:=true;
   }goto loc2_phil_2_C1;
   when !(left_fork(phil_2).owner==null && right_fork(phil_2).owner==null && phil_2.hungry)
   do invisible {} goto loc2_phil_2_N1;

   //philosopher 3
   when (left_fork(phil_3).owner==null && right_fork(phil_3).owner==null && phil_3.hungry)
   do invisible{
      left_fork(phil_3).owner:=phil_3;
      right_fork(phil_3).owner:=phil_3;
      phil_3.eating:=true;
   }goto loc2_phil_3_C1;
   when !(left_fork(phil_3).owner==null && right_fork(phil_3).owner==null && phil_3.hungry)
   do invisible {} goto loc2_phil_3_N1;

   //philosopher 4
   when (left_fork(phil_4).owner==null && right_fork(phil_4).owner==null && phil_4.hungry)
   do invisible{
      left_fork(phil_4).owner:=phil_4;
      right_fork(phil_4).owner:=phil_4;
      phil_4.eating:=true;
   }goto loc2_phil_4_C1;
   when !(left_fork(phil_4).owner==null && right_fork(phil_4).owner==null && phil_4.hungry)
   do invisible {} goto loc2_phil_4_N1;

   //philosopher 5
   when (left_fork(phil_5).owner==null && right_fork(phil_5).owner==null && phil_5.hungry)
   do invisible{
      left_fork(phil_5).owner:=phil_5;
      right_fork(phil_5).owner:=phil_5;
      phil_5.eating:=true;
   }goto loc2_phil_5_C1;
   when !(left_fork(phil_5).owner==null && right_fork(phil_5).owner==null && phil_5.hungry)
   do invisible {} goto loc2_phil_5_N1;

loc loc2_phil_1_C1:
   //changed={left_fork(phil_1).owner, right_fork(phil_1).owner, phil_1.eating}
   do invisible {} goto endloc;
loc loc2_phil_1_N1:
   //changed={}
   when (phil_1.eating && left_fork(phil_1).owner==phil_1 &&
   right_fork(phil_1).owner==phil_1 && !phil_1.hungry) do invisible {
      left_fork(phil_1).owner:=null;
      right_fork(phil_1).owner:=null;
      phil_1.eating:=false;
   }goto endloc;
   when !(phil_1.eating && left_fork(phil_1).owner==phil_1 &&
   right_fork(phil_1).owner==phil_1 && !phil_1.hungry) do invisible {} goto endloc;

loc loc2_phil_2_C1:
   //changed={left_fork(phil_2).owner, right_fork(phil_2).owner, phil_2.eating}
   do invisible {} goto endloc;
loc loc2_phil_2_N1:
   //changed={}
   when (phil_2.eating && left_fork(phil_2).owner==phil_2 &&
   right_fork(phil_2).owner==phil_2 && !phil_2.hungry) do invisible {
      left_fork(phil_2).owner:=null;
      right_fork(phil_2).owner:=null;
      phil_2.eating:=false;
   }goto endloc;
   when !(phil_2.eating && left_fork(phil_2).owner==phil_2 &&
```

```
                right_fork(phil_2).owner==phil_2 && !phil_2.hungry) do invisible {} goto endloc;

        loc loc2_phil_3_C1:
            //changed={left_fork(phil_3).owner, right_fork(phil_3).owner, phil_3.eating}
            do invisible {} goto endloc;
        loc loc2_phil_3_N1:
            //changed={}
            when (phil_3.eating && left_fork(phil_3).owner==phil_3 &&
            right_fork(phil_3).owner==phil_3 && !phil_3.hungry) do invisible {
                left_fork(phil_3).owner:=null;
                right_fork(phil_3).owner:=null;
                phil_3.eating:=false;
            }goto endloc;
            when !(phil_3.eating && left_fork(phil_3).owner==phil_3 &&
            right_fork(phil_3).owner==phil_3 && !phil_3.hungry) do invisible {} goto endloc;

        loc loc2_phil_4_C1:
            //changed={left_fork(phil_4).owner, right_fork(phil_4).owner, phil_4.eating}
            do invisible {} goto endloc;
        loc loc2_phil_4_N1:
            //changed={}
            when (phil_4.eating && left_fork(phil_4).owner==phil_4 &&
            right_fork(phil_4).owner==phil_4 && !phil_4.hungry) do invisible {
                left_fork(phil_4).owner:=null;
                right_fork(phil_4).owner:=null;
                phil_4.eating:=false;
            }goto endloc;
            when !(phil_4.eating && left_fork(phil_4).owner==phil_4 &&
            right_fork(phil_4).owner==phil_4 && !phil_4.hungry) do invisible {} goto endloc;

        loc loc2_phil_5_C1:
            //changed={left_fork(phil_5).owner, right_fork(phil_5).owner, phil_5.eating}
            do invisible {} goto endloc;
        loc loc2_phil_5_N1:
            //changed={}
            when (phil_5.eating && left_fork(phil_5).owner==phil_5 &&
            right_fork(phil_5).owner==phil_5 && !phil_5.hungry) do invisible {
                left_fork(phil_5).owner:=null;
                right_fork(phil_5).owner:=null;
                phil_5.eating:=false;
            }goto endloc;
            when !(phil_5.eating && left_fork(phil_5).owner==phil_5 &&
            right_fork(phil_5).owner==phil_5 && !phil_5.hungry) do invisible {} goto endloc;

        loc endloc:
            do {/*Bogor creates a state*/} goto loc1;
    }
}
```

**Fig. 19.** Dining philosophers problem: BIR model

Let's check the deadlock and correctness of the desired properties:

user@linux-cdhm:~> asm2bogor.sh diningPhilosophers.asm
Bogor v.1.2 (build <version>)
(c) Copyright by Kansas State University

Web: http://bogor.projects.cis.ksu.edu

Transitions: 8801, States: 353, Matched States: 4929, Max Depth: 525, Errors found: 0, Used Memory: 1MB
** diningPhilosophers is not in DEADLOCK
**LTLSPEC NAME ltl_HungryToEatingPhil1:=g((hungry(phil_1) and isUndef(owner(fork_1)) and isUndef(owner(fork_2)))
implies (eating(phil_1) iff (owner(fork_1)=phil_1 and owner(fork_2)=phil_1))) is true
**LTLSPEC NAME ltl_HungryToEatingPhil2:=g((hungry(phil_2) and isUndef(owner(fork_2)) and isUndef(owner(fork_3)))
implies (eating(phil_2) iff (owner(fork_1)=phil_2 and owner(fork_2)=phil_2))) is true
**LTLSPEC NAME ltl_HungryToEatingPhil3:=g((hungry(phil_3) and isUndef(owner(fork_3)) and isUndef(owner(fork_4)))
implies (eating(phil_3) iff (owner(fork_1)=phil_3 and owner(fork_2)=phil_3))) is true

**LTLSPEC NAME ltl_HungryToEatingPhil4:=g((hungry(phil_4) and isUndef(owner(fork_4)) and isUndef(owner(fork_5))) implies (eating(phil_4) iff (owner(fork_1)=phil_4 and owner(fork_2)=phil_4))) is true
**LTLSPEC NAME ltl_HungryToEatingPhil5:=g((hungry(phil_5) and isUndef(owner(fork_5)) and isUndef(owner(fork_1))) implies (eating(phil_5) iff (owner(fork_1)=phil_5 and owner(fork_2)=phil_5))) is true
**LTLSPEC NAME ltl_chkFork1:= g(not(owner(fork_1)=phil_1 and owner(fork_1)=phil_5)) is true
**LTLSPEC NAME ltl_chkFork2:= g(not(owner(fork_2)=phil_1 and owner(fork_2)=phil_2)) is true
**LTLSPEC NAME ltl_chkFork3:= g(not(owner(fork_3)=phil_2 and owner(fork_3)=phil_3)) is true
**LTLSPEC NAME ltl_chkFork4:= g(not(owner(fork_4)=phil_3 and owner(fork_4)=phil_4)) is true
**LTLSPEC NAME ltl_chkFork5:= g(not(owner(fork_5)=phil_4 and owner(fork_5)=phil_5)) is true
**LTLSPEC NAME  ltl_eatingOfNeighbours:=  g(not(eating(phil_1)  and  eating(phil_2))  or  not(eating(phil_2)  and eating(phil_3)) or not(eating(phil_3) and eating(phil_4)) or not(eating(phil_4) and eating(phil_5)) or not(eating(phil_5) and eating(phil_1))) is true
Done!

## 7.4.    Critical section problem

Critical  section  problem  is  a  fundamental  problem  of  concurrent  programming.  It contains two or more processes that execute concurrently and access the shared variables, data and resources. The access must be controlled otherwise some processes can obtain an inconsistent view of the shared variables, data and resources. Hence, each process executes the following code:

*while (true){*
> *entry-section*
> *critical section*
> *exit-section*
> *noncritical section*
*}*

The  entry-  and  exit-section  must  satisfy  the  exclusive  access  to  the  shared  resources: when a process is executing in its critical section, no other processes can be executing in their critical section. Binary semaphores are one of the synchronization mechanisms which are used to ensure exclusive use. It initialized to green and then changed through two operations,  named  *wait*  and  *signal*;  when  a  process  calls  the  wait  operation,  if  the semaphore is green, it can enter the critical section and changes the semaphore to red, otherwise  it  waits  until  the  semaphore  is  green;  and  when  a  process  calls  the  signal operation, it can exit from the critical section and changes the semaphore to green.

Fig. 20 is the AsmetaL model of the critical section problem.

```
asm criticalSectionProblem

import StandardLibrary
import LTLlibrary

signature:
    domain Process subsetof Agent
    enum domain BinSem = {GREEN | RED}
    enum domain ProcessStatus = {IDLE | ENTERING | CRITICAL | EXITING}
    dynamic controlled status: Process -> ProcessStatus
    dynamic controlled semaphore: BinSem
    dynamic monitored wantToEnter: Process -> Boolean
    dynamic monitored wantToExit: Process -> Boolean
    static process1: Process
    static process2: Process

definitions:
    rule r_wait =
        if (semaphore = GREEN) then
```

```
            par
                status(self) := CRITICAL
                semaphore := RED
            endpar
        endif

    rule r_signal =
        par
            status(self) := IDLE
            semaphore := GREEN
        endpar

    rule r_decideToEnter =
        if(status(self) = IDLE) then
            if(wantToEnter(self)) then
                status(self) := ENTERING
            endif
        endif

    rule r_entry_section =
        if(status(self) = ENTERING) then
            r_wait[]
        endif

    rule r_decideToExit =
        if(status(self) = CRITICAL) then
            if(wantToExit(self)) then
                status(self) := EXITING
            endif
        endif

    rule r_exit_section =
        if(status(self) = EXITING) then
            r_signal[]
        endif

    rule r_processRun =
        par
            r_decideToEnter[]
            r_entry_section[]
            r_decideToExit[]
            r_exit_section[]
        endpar

    //if a process want to enter the critical section it must take the semaphore
    LTLSPEC NAME ltl_enterCriticalP1:= g((status(process1)=ENTERING and semaphore=GREEN) implies
                                        (status(process1)=CRITICAL iff semaphore=RED))
    LTLSPEC NAME ltl_enterCriticalP2:= g((status(process1)=ENTERING and semaphore=GREEN) implies
                                        (status(process1)=CRITICAL iff semaphore=RED))

    //the two processes can never be in the critical section at the same time
    LTLSPEC NAME ltl_MutualExclusion:= g(not(status(process1)=CRITICAL and
                                        status(process2)=CRITICAL))

    main rule r_Main = choose $p in Process with true do program($p)

default init s0:
    function status($p in Process) = IDLE
    function semaphore = GREEN
    agent Process: r_processRun[]
```

**Fig. 20.** Critical section problem: AsmetaL model

In the model, we consider the case where there are two processes, *process1* and *process2*. Each process has a controlled function, *status*, which records the process status and two monitored functions, *wantToEnter* and *wantToExit* which represent the process want to enter or exit from its critical section; The *wantToEnter(process1)* location, for example, is true if process1 want to enter the critical section.

To model the binary semaphores, we define the controlled function *binsem*. The behaviour of the wait and signal operations are simulated with the two rules, *r_wait* and *r_signal*.

We also declare some properties we want to check.

Safety properties

*LTLSPEC NAME ltl_enterCriticalP1:= g((status(process1)=ENTERING and semaphore=GREEN)*
*implies (status(process1)=CRITICAL iff semaphore=RED))*
*LTLSPEC NAME ltl_enterCriticalP2:= g((status(process1)=ENTERING and semaphore=GREEN)*
*implies (status(process1)=CRITICAL iff semaphore=RED))*

check that, always if *process$_i$* is in entry section (*status(process$_i$)=ENTERING*) and the semaphore is free (*semaphore=GREEN*), then it can enter the critical section if and only if it take the semaphore.

The safety property

*LTLSPEC NAME ltl_MutualExclusion:=g(not(status(process1)=CRITICAL and status(process2)=CRITICAL))*

checks that, the solution guarantees mutual exclusion.

Fig. 21 is the BIR code obtained from the mapping of the AsmetaL code represented in Fig. 20.

```
system criticalSectionProblem {

    fun ltl_enterCriticalP1() returns boolean =
        LTL.temporalProperty(
            Property.createObservableDictionary(
                Property.createObservableKey("P1", process1.status==ProcessStatus.ENTERING),
                Property.createObservableKey("P2", semaphore==BinSem.GREEN),
                Property.createObservableKey("P3", process1.status==ProcessStatus.CRITICAL),
                Property.createObservableKey("P4", semaphore==BinSem.RED)),
            LTL.always(LTL.implication(LTL.conjunction(LTL.prop("P1"),LTL.prop("P2")),
                                       LTL.equivalence(LTL.prop("P3"),LTL.prop("P4")))))
        );
    fun ltl_enterCriticalP2() returns boolean =
        LTL.temporalProperty(
            Property.createObservableDictionary(
                Property.createObservableKey("P1", process2.status==ProcessStatus.ENTERING),
                Property.createObservableKey("P2", semaphore==BinSem.GREEN),
                Property.createObservableKey("P3", process2.status==ProcessStatus.CRITICAL),
                Property.createObservableKey("P4", semaphore==BinSem.RED)),
            LTL.always(LTL.implication(LTL.conjunction(LTL.prop("P1"),LTL.prop("P2")),
                                       LTL.equivalence(LTL.prop("P3"),LTL.prop("P4")))))
        );
    fun ltl_MutualExclusion() returns boolean =
        LTL.temporalProperty(
            Property.createObservableDictionary(
                Property.createObservableKey("P1", process1.status==ProcessStatus.CRITICAL),
                Property.createObservableKey("P2", process2.status==ProcessStatus.CRITICAL)),
            LTL.always(LTL.negation(LTL.conjunction(.prop("P1"),LTL.prop("P2")))))
        );

    record Process {
        ProcessStatus status;//controlled
        boolean wantToEnter;//monitored
        boolean wantToExit;//monitored
    }
    enum BinSem {GREEN , RED}
    enum ProcessStatus {IDLE , ENTERING , CRITICAL , EXITING}
    BinSem semaphore;//controlled
    Process process1;//static
    Process process2;//static

    thread wantToEnter_process1_monitored() {
```

```
    loc loc0:
        do {process1.wantToEnter:=true;} goto loc0;
        do {process1.wantToEnter:=false;} goto loc0;
}
thread wantToExit_process1_monitored() {
    loc loc0:
        do {process1.wantToExit:=true;} goto loc0;
        do {process1.wantToExit:=false;} goto loc0;
}
thread wantToEnter_process2_monitored() {
    loc loc0:
        do {process2.wantToEnter:=true;} goto loc0;
        do {process2.wantToEnter:=false;} goto loc0;
}
thread wantToExit_process2_monitored() {
    loc loc0:
        do {process2.wantToExit:=true;} goto loc0;
        do {process2.wantToExit:=false;} goto loc0;
}

active thread MAIN() {
    loc loc0://initialization
        do {
            process1 := new Process;
            process2 := new Process;
            process1.status := ProcessStatus.IDLE;
            process2.status := ProcessStatus.IDLE;
            semaphore := BinSem.GREEN;

            //monitored functions
            start wantToEnter_process1_monitored();
            start wantToExit_process1_monitored();
            start wantToEnter_process2_monitored();
            start wantToExit_process2_monitored();
        }goto loc1;

    //rules
    loc loc1:
        //process 1
        when (process1.status==ProcessStatus.IDLE && process1.wantToEnter) do invisible {
            process1.status := ProcessStatus.ENTERING;
        }goto loc2_C1;
        when !(process1.status==ProcessStatus.IDLE && process1.wantToEnter) do invisible
        {} goto loc2_process1_N1;
        //process 2
        when (process2.status==ProcessStatus.IDLE && process2.wantToEnter) do invisible {
            process2.status := ProcessStatus.ENTERING;
        }goto loc2_C1;
        when !(process2.status==ProcessStatus.IDLE && process2.wantToEnter) do invisible
        {} goto loc2_process2_N1;

    loc loc2_C1:
        do invisible{}goto endloc;

    loc loc2_process1_N1:
        when (process1.status==ProcessStatus.ENTERING && semaphore==BinSem.GREEN) do invisible {
            process1.status := ProcessStatus.CRITICAL;
            semaphore := BinSem.RED;
        }goto loc2_C1;
        when !(process1.status==ProcessStatus.ENTERING && semaphore==BinSem.GREEN) do invisible
        {} goto loc2_process1_N1_N2;

    loc loc2_process1_N1_N2:
        when (process1.status==ProcessStatus.CRITICAL && process1.wantToExit) do invisible {
            process1.status := ProcessStatus.EXITING;
        }goto loc2_C1;
        when !(process1.status==ProcessStatus.CRITICAL && process1.wantToExit) do invisible
        {} goto loc2_process1_N1_N2_N3;

    loc loc2_process1_N1_N2_N3:
        when (process1.status == ProcessStatus.EXITING) do invisible {
```

```
                process1.status := ProcessStatus.IDLE;
                semaphore := BinSem.GREEN;
            }goto endloc;
            when !(process1.status == ProcessStatus.EXITING) do invisible
            {} goto endloc;

        loc loc2_process2_N1:
            when (process2.status==ProcessStatus.ENTERING && semaphore==BinSem.GREEN) do invisible {
                process2.status := ProcessStatus.CRITICAL;
                semaphore := BinSem.RED;
            }goto loc2_C1;
            when !(process2.status==ProcessStatus.ENTERING && semaphore==BinSem.GREEN) do invisible
            {} goto loc2_process2_N1_N2;

        loc loc2_process2_N1_N2:
            when (process2.status==ProcessStatus.CRITICAL && process2.wantToExit) do invisible {
                process2.status := ProcessStatus.EXITING;
            }goto loc2_C1;
            when !(process2.status==ProcessStatus.CRITICAL && process2.wantToExit) do invisible
            {} goto loc2_process2_N1_N2_N3;

        loc loc2_process2_N1_N2_N3:
            when (process2.status==ProcessStatus.EXITING) do invisible {
                process2.status := ProcessStatus.IDLE;
                semaphore := BinSem.GREEN;
            }goto endloc;
            when !(process2.status==ProcessStatus.EXITING) do invisible
            {} goto endloc;

        loc endloc:
            do {/*Bogor creates a state*/} goto loc1;
    }
}
```

**Fig. 21.** Critical section problem: BIR model

Let's check the deadlock and correctness of the property:

```
user@linux-cdhm:~> asm2bogor.sh criticalSectionProblem.asm
Bogor v.1.2 (build <version>)
(c) Copyright by Kansas State University

Web: http://bogor.projects.cis.ksu.edu

Transitions: 3265, States: 193, Matched States: 1729, Max Depth: 443, Errors found: 0, Used Memory: 0MB
**criticalSectionProblem is not in DEADLOCK
**LTLSPEC   NAME   ltl_enterCriticalP1:=   g((status(process1)=ENTERING   and   semaphore=GREEN)   implies
(status(process1)=CRITICAL iff semaphore=RED)) is true
**LTLSPEC   NAME   ltl_enterCriticalP2:=   g((status(process1)=ENTERING   and   semaphore=GREEN)   implies
(status(process1)=CRITICAL iff semaphore=RED)) is true
**LTLSPEC NAME ltl_MutualExclusion:= g(not(status(process1)=CRITICAL and status(process2)=CRITICAL)) is true
Done!
```

## 7.5. Tic Tac Toe simulator

Tic Tac Toe simulator is a game played on a 3×3 square board [29]. It lets to play the game with the computer. The symbol of the use is cross (i.e., '*X*') and the computer uses of nought (i.e., '*O*'). At each step, the user and the computer, make a move in turn by turns. The user always goes first. They put their symbols in one of the empty cells until a player wins or all cells are filled. A player wins if there is a row, column or diagonal in which all the cells contain her symbol, and if there is no empty cell and neither player wins, the game is a tie.

Fig. 22 is the AsmetaL model.

```
asm ticTacToe_simulator

import StandardLibrary
import LTLlibrary

signature:
    abstract domain Cell
    enum domain SignDomain = {CROSS | NOUGHT | EMPTY}
    enum domain StatusDomain = {TURN_USER | TURN_PC}
    dynamic controlled sign: Cell -> SignDomain
    static board_1_1: Cell
    static board_1_2: Cell
    static board_1_3: Cell
    static board_2_1: Cell
    static board_2_2: Cell
    static board_2_3: Cell
    static board_3_1: Cell
    static board_3_2: Cell
    static board_3_3: Cell
    controlled status: StatusDomain
    monitored userChoiceCell: Cell
    derived winner: SignDomain -> Boolean
    derived endOfGame: Boolean

definitions:
    function winner($s in SignDomain) =
        //rows
        ((sign(board_1_1)=$s and sign(board_1_2)=$s and sign(board_1_3)=$s) or
         (sign(board_2_1)=$s and sign(board_2_2)=$s and sign(board_2_3)=$s) or
         (sign(board_3_1)=$s and sign(board_3_2)=$s and sign(board_3_3)=$s))or
        //cols
        ((sign(board_1_1)=$s and sign(board_2_1)=$s and sign(board_3_1)=$s) or
         (sign(board_1_2)=$s and sign(board_2_2)=$s and sign(board_3_2)=$s) or
         (sign(board_1_3)=$s and sign(board_2_3)=$s and sign(board_3_3)=$s))or
        //main chord
        (sign(board_1_1)=$s and sign(board_2_2)=$s and sign(board_3_3)=$s)  or
        //secondary chord
        (sign(board_1_3)=$s and sign(board_2_2)=$s and sign(board_3_1)=$s)

    function endOfGame =
        winner(CROSS) or
        winner(NOUGHT) or
        (forall $r in Cell with sign($r)!=EMPTY)

    rule r_moveUser =
        let ($x = userChoiceCell) in
            if(sign($x)=EMPTY) then
                par
                    sign($x) := CROSS
                    status := TURN_PC
                endpar
            endif
        endlet

    rule r_movePC =
        choose $x in Cell with sign($x)=EMPTY do
            par
                sign($x) :=  NOUGHT
                status := TURN_USER
            endpar

    //safety property: checks the end of the game: which player can win?
    LTLSPEC NAME ltl_endOfGame:= g(endOfGame implies (winner(CROSS) or winner(NOUGHT) or
                                       (not(winner(NOUGHT)) and not(winner(CROSS)))))

    //safety property: the user and the computer cannot win together
    LTLSPEC NAME ltl_notBothWin:= g(not(winner(CROSS) and winner(NOUGHT)))
```

```
   main rule r_Main =
      if(not(endOfGame)) then
         if(status = TURN_USER) then
            r_moveUser[]
         else
            r_movePC[]
         endif
      endif

default init s0:
   function status = TURN_USER
   function sign($c in Cell) = EMPTY
```

**Fig. 21.** Tic Tac Toe simulator: AsmetaL model

Each cell of the game board has a sign, i.e., cross, nought or empty. To represent the cells, we define the abstract domain, *Cell*, consists of the controlled function *sign*. The game board is created by the nine static variable *board_i_j* where 1≤i≤3 and 1≤j≤3. The turn of the user and the computer is showed by the controlled function, *status*. Moreover, the move of the user is specified through the monitored location *userChoiceCell*.

We also define two derived function to control the state of the system: the *winner($s in SignDomain)* function, checks all the horizontal, vertical and diagonal rows and returns true if the player that her symbol is *$s*, wins the game; and the *endOfGame* shows the game is ended.

The two macro rules simply simulate the behaviour of the players:

- In the *r_moveUser* rule, the user chooses a cell through the monitored location *userChoiceCell* and if it is empty, then he places a cross on the cell and changes the turn.
- In the *r_movePC* rule, the computer non-deterministically chooses an empty cell, fills it and changes the turn.

In the main rule, if the game is not ended, we check the turn and call the proper rule. At the beginning the board is empty and the user must go first (i.e., *status=TURN_USER*).

Let's see some properties which help us to check the correctness of the model.

Safety property

*LTLSPEC NAME ltl_endOfGame:= g(endOfGame implies (winner(CROSS) or*
                          *winner(NOUGHT) or*
                          *(not(winner(NOUGHT)) and not(winner(CROSS)))))*

says that, always the user or computer can win the game, or the result of the game will in a tie.

The safety property

*LTLSPEC NAME ltl_notBothWin:= g(not(winner(CROSS) and winner(NOUGHT)))*

checks that, the user and computer cannot win together.

Fig. 22 is the equivalent BIR code of the AsmetaL model represented in Fig. 21.

```
system ticTacToe_simulator {

    fun ltl_endOfGame() returns boolean =
       LTL.temporalProperty(
          Property.createObservableDictionary(
             Property.createObservableKey("P1", endOfGame()),
             Property.createObservableKey("P2", winner(Sign.CROSS)),
             Property.createObservableKey("P3", winner(Sign.NOUGHT))),
```

```
        LTL.always(LTL.implication(
                    LTL.prop("P1"),
                LTL.disjunction(LTL.disjunction(LTL.prop("P2"),LTL.prop("P3")),
                                    LTL.conjunction(LTL.negation(LTL.prop("P2")),
                                                    LTL.negation(LTL.prop("P3"))))))
    );
fun ltl_notBothWin() returns boolean =
    LTL.temporalProperty(
        Property.createObservableDictionary(
            Property.createObservableKey("P1", winner(Sign.CROSS)),
            Property.createObservableKey("P2", winner(Sign.NOUGHT))),
        LTL.always(LTL.negation(LTL.conjunction(LTL.prop("P1"),LTL.prop("P2")))))
    );

record Cell {
    Sign sign;//controlled
}
typealias Coord int;
enum Sign {CROSS, NOUGHT, EMPTY}
enum Status {TURN_USER, TURN_PC}
Cell board_1_1;//static
Cell board_1_2;//static
Cell board_1_3;//static
Cell board_2_1;//static
Cell board_2_2;//static
Cell board_2_3;//static
Cell board_3_1;//static
Cell board_3_2;//static
Cell board_3_3;//static
Status status;//controlled
Cell userChoiceCell;//monitored

thread userChoiceCell_monitored() {
    loc loc0 :
        do {userChoiceCell:=board_1_1;} goto loc0;
        do {userChoiceCell:=board_1_2;} goto loc0;
        do {userChoiceCell:=board_1_3;} goto loc0;
        do {userChoiceCell:=board_2_1;} goto loc0;
        do {userChoiceCell:=board_2_2;} goto loc0;
        do {userChoiceCell:=board_2_3;} goto loc0;
        do {userChoiceCell:=board_3_1;} goto loc0;
        do {userChoiceCell:=board_3_2;} goto loc0;
        do {userChoiceCell:=board_3_3;} goto loc0;
}

fun winner(Sign s) returns boolean =
    (//rows
        ((board_1_1.sign==s && board_1_2.sign==s && board_1_3.sign==s) ||
         (board_2_1.sign==s && board_2_2.sign==s && board_2_3.sign==s) ||
         (board_3_1.sign==s && board_3_2.sign==s && board_3_3.sign==s)) ||
    //cols
        ((board_1_1.sign==s && board_2_1.sign==s && board_3_1.sign==s) ||
         (board_1_2.sign==s && board_2_2.sign==s && board_3_2.sign==s) ||
         (board_1_3.sign==s && board_2_3.sign==s && board_3_3.sign==s)) ||
    //main chord
        (board_1_1.sign==s && board_2_2.sign==s && board_3_3.sign==s) ||
    //secondary chord
        (board_1_3.sign==s && board_2_2.sign==s && board_3_1.sign==s));

fun endOfGame() returns boolean =
    ((winner(Sign.CROSS) || winner(Sign.NOUGHT)) ||
     (board_1_1.sign!=Sign.EMPTY && board_1_2.sign!=Sign.EMPTY && board_1_3.sign!=Sign.EMPTY &&
      board_2_1.sign!=Sign.EMPTY && board_2_2.sign!=Sign.EMPTY && board_2_3.sign!=Sign.EMPTY &&
      board_3_1.sign!=Sign.EMPTY && board_3_2.sign!=Sign.EMPTY && board_3_3.sign!=Sign.EMPTY));

active thread MAIN() {
    loc loc0://initialization
        do {
            board_1_1 := new Cell;
            board_1_2 := new Cell;
            board_1_3 := new Cell;
```

```
        board_2_1 := new Cell;
        board_2_2 := new Cell;
        board_2_3 := new Cell;
        board_3_1 := new Cell;
        board_3_2 := new Cell;
        board_3_3 := new Cell;
        userChoiceCell := new Cell;
        board_1_1.sign := Sign.EMPTY;
        board_1_2.sign := Sign.EMPTY;
        board_1_3.sign := Sign.EMPTY;
        board_2_1.sign := Sign.EMPTY;
        board_2_2.sign := Sign.EMPTY;
        board_2_3.sign := Sign.EMPTY;
        board_3_1.sign := Sign.EMPTY;
        board_3_2.sign := Sign.EMPTY;
        board_3_3.sign := Sign.EMPTY;
        status := Status.TURN_USER;

        //monitored functions
        start userChoiceCell_monitored();
    }goto loc1;

//rules
loc loc1:
    when (!endOfGame() && status==Status.TURN_USER && userChoiceCell.sign==Sign.EMPTY)
    do invisible {
        userChoiceCell.sign := Sign.CROSS;
        status := Status.TURN_PC;
    }goto endloc;

    when (!endOfGame() && !(status==Status.TURN_USER) && board_1_1.sign==Sign.EMPTY)
    do invisible {
        board_1_1.sign := Sign.NOUGHT;
        status := Status.TURN_USER;
    }goto endloc;

    when (!endOfGame() && !(status==Status.TURN_USER) && board_1_2.sign==Sign.EMPTY)
    do invisible {
        board_1_2.sign := Sign.NOUGHT;
        status := Status.TURN_USER;
    }goto endloc;

    when (!endOfGame() && !(status==Status.TURN_USER) && board_1_3.sign==Sign.EMPTY)
    do invisible {
        board_1_3.sign := Sign.NOUGHT;
        status := Status.TURN_USER;
    }goto endloc;

    when (!endOfGame() && !(status==Status.TURN_USER) && board_2_1.sign==Sign.EMPTY)
    do invisible {
        board_2_1.sign := Sign.NOUGHT;
        status := Status.TURN_USER;
    }goto endloc;

    when (!endOfGame() && !(status==Status.TURN_USER) && board_2_2.sign==Sign.EMPTY)
    do invisible {
        board_2_2.sign := Sign.NOUGHT;
        status := Status.TURN_USER;
    }goto endloc;

    when (!endOfGame() && !(status==Status.TURN_USER) && board_2_3.sign==Sign.EMPTY)
    do invisible {
        board_2_3.sign := Sign.NOUGHT;
        status := Status.TURN_USER;
    }goto endloc;

    when (!endOfGame() && !(status==Status.TURN_USER) && board_3_1.sign==Sign.EMPTY)
    do invisible {
        board_3_1.sign := Sign.NOUGHT;
        status := Status.TURN_USER;
    }goto endloc;
```

```
        when (!endOfGame() && !(status==Status.TURN_USER) && board_3_2.sign==Sign.EMPTY)
        do invisible {
            board_3_2.sign := Sign.NOUGHT;
            status := Status.TURN_USER;
        }goto endloc;

        when (!endOfGame() && !(status==Status.TURN_USER) && board_3_3.sign==Sign.EMPTY)
        do invisible {
            board_3_3.sign := Sign.NOUGHT;
            status := Status.TURN_USER;
        }goto endloc;

    loc endloc:
        do {/*Bogor creates a state*/} goto loc1;
    }
}
```

**Fig. 22.** Tic Tac Toe simulator: BIR model

Let's see the execution of BIR code in order to check the deadlock and correctness of the properties:

user@linux-cdhm:~> asm2bogor.sh ticTacToe_simulator.asm
Bogor v.1.2 (build <version>)
(c) Copyright by Kansas State University

Web: http://bogor.projects.cis.ksu.edu

Transitions: 596638, States: 49304, Matched States: 470880, Max Depth: 57, Errors found: 0, Used Memory: 9MB
** ticTacToe_simulator is not in DEADLOCK
**LTLSPEC NAME ltl_endOfGame:= g(endOfGame implies (winner(CROSS) or winner(NOUGHT) or (not(winner(NOUGHT)) and not(winner(CROSS))))) is true
**LTLSPEC NAME ltl_notBothWin:= g(not(winner(CROSS) and winner(NOUGHT))) is true
Done!

### 7.6. Ferryman simulator

Ferryman simulator is a river crossing puzzle in which a ferryman must transport a goat, wolf and cabbage safety from the left bank to the right bank of the river using a boat. The boat is only big enough to carry two actors at a time. Hence, the ferryman can cross the river alone or with one of the actors. Moreover, since the wolf can kill the goat and the goat can eat the cabbage, the ferryman cannot leave alone the goat with the cabbage and wolf. At the beginning all the actors are on the left side of the river.

Fig. 23 is the AsmetaL model which implements a correct behaviour of the ferryman, hence, no meal situations can happen.

```
asm ferryman

import StandardLibrary
import LTLlibrary

signature:
    abstract domain Actors
    enum domain SideDomain = {LEFT | RIGHT}
    dynamic controlled position: Actors -> SideDomain
    static goodCouple: Prod(Actors, Actors) -> Boolean
    derived goodSituationSide: Prod(Actors, Actors, SideDomain) -> Boolean
    derived allOnRightSide: Boolean
    static ferryman: Actors
    static goat: Actors
    static cabbage: Actors
```

```
      static wolf: Actors

definitions:
    function goodCouple($a in Actors, $b in Actors) =
        ($a=goat implies ($b!=cabbage and $b!=wolf)) and
        ($b=goat implies ($a!=cabbage and $a!=wolf))

    function goodSituationSide($a in Actors, $b in Actors, $s in SideDomain) =
        (position($a)=$s and position($b)=$s) implies goodCouple($a, $b)

    function allOnRightSide =
        (forall $a in Actors with position($a) = RIGHT)

    rule r_travelLeftToRight =
        if(position(ferryman)= LEFT) then
            choose $a in Actors with $a!=ferryman and position($a) = LEFT and
                    (forall $x in Actors, $y in Actors with (($x!=$a and $y!=$a) implies
                                                    goodSituationSide($x, $y, LEFT))) do
                par
                    position(ferryman) := RIGHT
                    position($a) := RIGHT
                endpar
            ifnone
                position(ferryman) := RIGHT
            endif

    rule r_travelRightToLeft =
        if(position(ferryman)=RIGHT and not(allOnRightSide)) then
            choose $c in Actors with $c!=ferryman and position($c) = RIGHT
                            and (exist $k in Actors with not(goodSituationSide($c, $k, RIGHT))) do
                par
                    position(ferryman) := LEFT
                    position($c) := LEFT
                endpar
            ifnone
                position(ferryman) := LEFT
            endif

    //safety properties: check the security
    LTLSPEC NAME ltl_cabbageIsSecure:=  g(position(goat)=position(cabbage)   implies
                                            position(goat)=position(ferryman))
    LTLSPEC NAME ltl_goatIsSecure:= g(position(wolf)=position(goat)    implies
                                            position(wolf)=position(ferryman))

    //counter-example: Is there a solution?
    LTLSPEC NAME ltl_noSolution:= g(not(allOnRightSide))

    main rule r_Main =
        par
            r_travelLeftToRight[]
            r_travelRightToLeft[]
        endpar

default init s0:
    function position($a in Actors) = LEFT
```

**Fig. 23.** Ferryman simulator: AsmetaL model

To represent the two bank of the river we define the enum domain *SideDomain*. Moreover, the agents are modelled by the abstract domain *Actors* consists of the controlled location *position* which shows the agent is on the left or the right bank.

We define three boolean derived functions which can check the situation of the puzzle:

- *goodCouple($a in Actors, $b in Actors)* checks that the two actors $a and $b are compatible.

- *goodSituationSide($a in Actors, $b in Actors, $s in SideDomain)* checks that the two actors $a and $b can stay on the same bank alone just if they are a good couple.
- *allOnRightSide* is true if all the actors are on the right bank of the river.

Let's see the main rule which of two rules *r_travelLeftToRight* and *r_travelRightToLeft*. These rules model the behaviour of the system: if the ferryman is on left bank, he can choose an actor such that all the actors that remains on the left bank are good couples, then he crosses the river alone or with the chosen actor; and when the ferryman is on the right bank and an actor exists on the left bank, he goes back to the left bank and if there is an actor who cannot stay alone with another one on the right bank, he carries him with himself.

To check the correctness of the model we have defined some properties.

First of all, we want to check the security. Safety properties

*LTLSPEC NAME ltl_cabbageIsSecure:= g(position(goat)=position(cabbage) implies*
*position(goat)=position(ferryman))*

*LTLSPEC NAME ltl_goatIsSecure:= g(position(wolf)=position(goat) implies*
*position(wolf)=position(ferryman))*

check that, the ferryman is on the bank where the goat and the cabbage are or the wolf and the goat are.

Safety property

*LTLSPEC NAME ltl_noSolution:= g(not(allOnRightSide))*

says that, globally there is no state in which all the actors are on the right bank of the river. If it is not satisfied, Bogor generates a counter-example which is the solution of the ferryman puzzle.

Fig. 24 is the BIR code obtained from the mapping of the AsmetaL code represented in Fig. 23. Since the BIR code is too big, we do not show the complete code.

```
system ferryman {

    fun ltl_cabbageIsSecure() returns boolean =
        LTL.temporalProperty(
            Property.createObservableDictionary(
                Property.createObservableKey("P1", goat.position==cabbage.position),
                Property.createObservableKey("P2", goat.position==ferryman.position)),
            LTL.always(LTL.implication(LTL.prop("P1"),LTL.prop("P2")))
        );
    fun ltl_goatIsSecure() returns boolean =
        LTL.temporalProperty(
            Property.createObservableDictionary(
                Property.createObservableKey("P1", wolf.position==goat.position),
                Property.createObservableKey("P2", wolf.position==ferryman.position)),
            LTL.always(LTL.implication(LTL.prop("P1"),LTL.prop("P2")))
        );
    fun ltl_noSolution() returns boolean =
        LTL.temporalProperty(
            Property.createObservableDictionary(
                Property.createObservableKey("P1", allOnRightSide())),
            LTL.always(LTL.negation(LTL.prop("P1")))
        );

    record Actors {
        SideDomain position;//controlled
    }
    enum SideDomain {LEFT, RIGHT}
```

```
Actors ferryman;//static
Actors goat;//static
Actors cabbage;//static
Actors wolf;//static

fun goodCouple(Actors a, Actors b) returns boolean = //p->q = ~pvq
    ((((!(a==goat)) ||(b!=wolf && b!=cabbage)) &&
     ((!(b==goat)) ||(a!=wolf && a!=cabbage)) ? true : false);
fun goodSituationSide(Actors a, Actors b, SideDomain s) returns boolean =
    ((((!(a.position==s && b.position==s))||(goodCouple(a,b))) ? true : false);
fun allOnRightSide() returns boolean =
    ((ferryman.position==SideDomain.RIGHT && goat.position==SideDomain.RIGHT &&
     cabbage.position==SideDomain.RIGHT && wolf.position==SideDomain.RIGHT )? true : false);

active thread MAIN() {
    loc loc0://initialization
        do {
            ferryman:= new Actors;
            goat:= new Actors;
            cabbage:= new Actors;
            wolf:= new Actors;
            ferryman.position:=SideDomain.LEFT;
            goat.position:=SideDomain.LEFT;
            cabbage.position:=SideDomain.LEFT;
            wolf.position:=SideDomain.LEFT;
        }goto loc1;

    //rules
    loc loc1:
        when (ferryman.position==SideDomain.LEFT &&
            goat!=ferryman && goat.position==SideDomain.LEFT &&
            ((((!(ferryman!=goat && ferryman!=goat)) ||
            (goodSituationSide(ferryman,ferryman,SideDomain.LEFT))) &&
            ((!(ferryman!=goat && cabbage!=goat)) ||
            (goodSituationSide(ferryman,cabbage,SideDomain.LEFT))) &&
            ((!(ferryman!=goat && wolf!=goat)) ||
            (goodSituationSide(ferryman,wolf,SideDomain.LEFT))) &&
            ((!(cabbage!=goat && ferryman!=goat)) ||
            (goodSituationSide(cabbage,ferryman,SideDomain.LEFT))) &&
            ((!(cabbage!=goat && cabbage!=goat)) ||
            (goodSituationSide(cabbage,cabbage,SideDomain.LEFT))) &&
            ((!(cabbage!=goat && wolf!=goat)) ||
            (goodSituationSide(cabbage,wolf,SideDomain.LEFT))) &&
            ((!(wolf!=goat && ferryman!=goat)) ||
            (goodSituationSide(wolf,ferryman,SideDomain.LEFT))) &&
            ((!(wolf!=goat && cabbage!=goat)) ||
            (goodSituationSide(wolf,cabbage,SideDomain.LEFT))) &&
            ((!(wolf!=goat && wolf!=goat)) ||
            (goodSituationSide(wolf,wolf,SideDomain.LEFT))))) do invisible {
                ferryman.position := SideDomain.RIGHT;
                goat.position := SideDomain.RIGHT;
            }goto loc2_C1;
            .
            .
            .
    loc endloc:
        do {/*Bogor creates a state*/} goto loc1;
    }
}
```

**Fig. 24.** Ferryman simulator: BIR model

Let's check the deadlock and correctness of the properties:

user@linux-cdhm:~> asm2bogor.sh ferryman.asm
Bogor v.1.2 (build <version>)
(c) Copyright by Kansas State University

Web: http://bogor.projects.cis.ksu.edu

Transitions: 63, States: 11, Matched States: 13, Max Depth: 25, Errors found: 1, Used Memory: 0MB
** ferryman is not in DEADLOCK
**LTLSPEC NAME ltl_cabbageIsSecure:= g(position(goat)=position(cabbage)   implies position(goat)=position(ferryman))
is true
**LTLSPEC NAME ltl_goatIsSecure:= g(position(wolf)=position(goat)   implies position(wolf)=position(ferryman)) is true
**LTLSPEC NAME ltl_noSolution:= g(not(allOnRightSide)) is false
Generating error trace O...
Done!

As we prospect, the ltl_noSolution property is not satisfied. This means that, there is a solution for the ferryman puzzle. Fig. 25 shows the counter-example. We use the Bogor which comes wrapped as a plug in for *Eclipse* [30]. It provides a graphical user interface and a variety of visualization facilities for the BIR models.

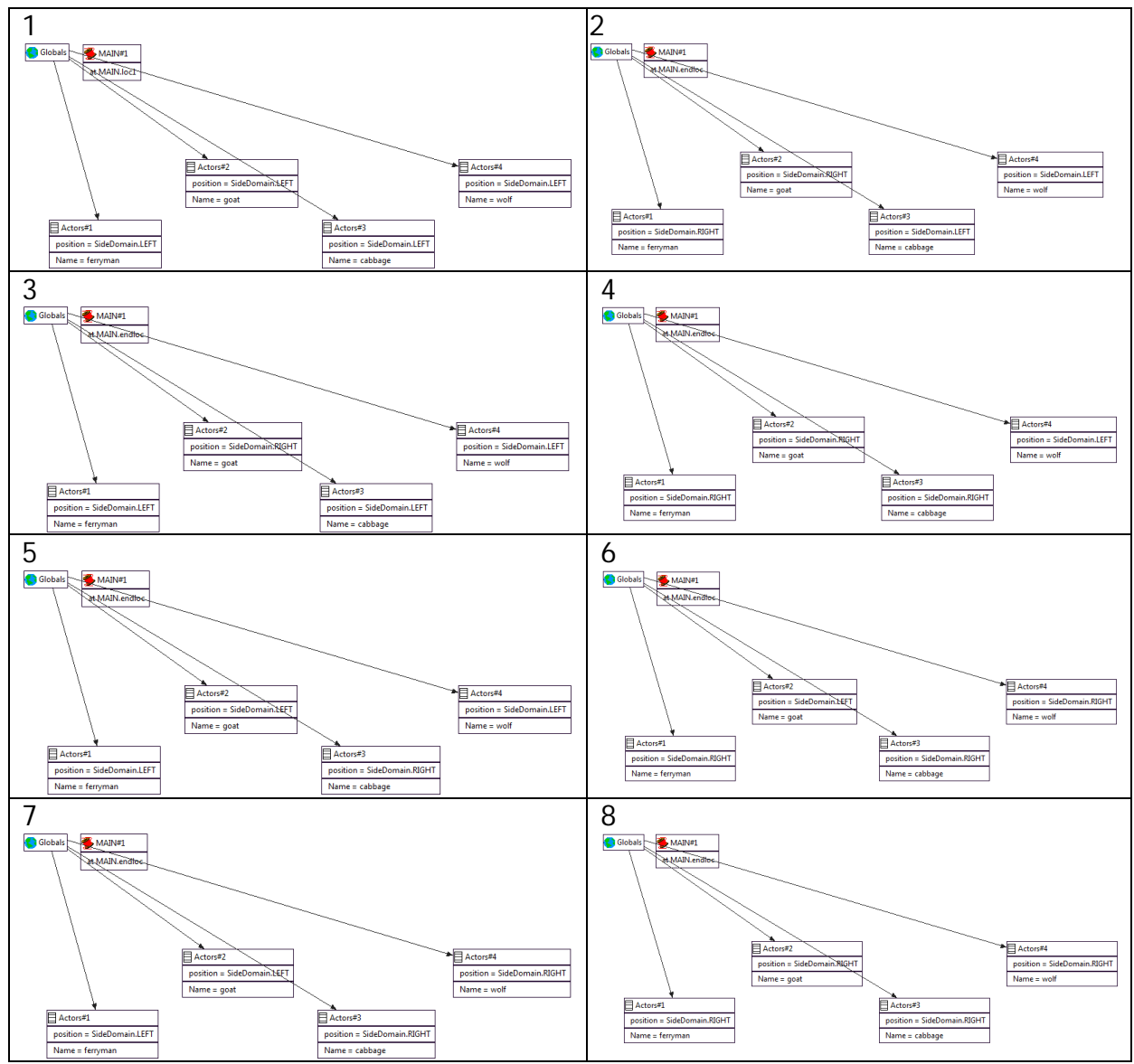

**Fig. 25.** Bogor counter example: heap structure displays

## 8. Conclusions

ASM is a suitable modeling language to specify different software systems like pub-sub and event-based systems. AsmetaL is a meta-model based language to represent ASM specifications. However, a proper modeling language should be equipped with the analysis capabilities to let the designers discover different properties on it. Hence, in this paper we present an approach for the verification of Asmetal specification. To do so, we propose an implementation framework to automatically encode AsmetaL specifications along with the designed properties to BIR. Then, Bogor evaluates the model through mode checking. Our approach has some key characteristics which make it suitable for analysis: (1) automation; designers can perform the verification through the "push button" mechanism, (2) designers do not need to understand Bogor, and (3) most of the constructs and statements in the AsmetaL are covered by our approach. Limitations are due to the model checker restriction over infinite domains.

## Acknowledgment

I would like to thank Elvinia Riccobene and Paolo Arcaini for their insightful comments on earlier versions of the paper.